\documentclass[usenatbib]{mn2e}

\usepackage{graphicx}
\usepackage{times}
\usepackage{amssymb}
\usepackage{amsmath}
\title[Radiation thrust in Schwarzschild space-time]
      {Particles under radiation thrust in Schwarzschild space-time from
       a flux perpendicular to the equatorial plane}
\author[D. Bini, A. Geralico, R. T. Jantzen, O. Semer\'ak]
       {D. Bini,$^{1,2,3}$
        A. Geralico,$^{1,2}$
        R. T. Jantzen$^{4,3}$ and
        O. Semer\'ak$^{5}$\\
        $^1$Istituto per le Applicazioni del Calcolo ``M. Picone", CNR I-00185 Rome, Italy\\
        $^2$International Center for Relativistic Astrophysics -- I.C.R.A.,
            University of Rome ``La Sapienza", I-00185 Rome, Italy\\
        $^3$INFN sezione di Napoli, Complesso Universitario di Monte S. Angelo,
            Via Cintia, Edificio 6, 80126 Napoli, Italy\\
        $^4$Department of Mathematics and Statistics, Villanova University, Villanova, PA 19085, USA\\
        $^5$Institute of Theoretical Physics, Faculty of Mathematics and Physics,
            Charles University in Prague, Czech Republic}

\begin{document}

\date{}

\pagerange{\pageref{firstpage}--\pageref{lastpage}} \pubyear{}

\maketitle

\label{firstpage}

\begin{abstract}
Motivated by the picture of a thin accretion disc around a black hole, radiating mainly in the direction perpendicular to its plane, we study the motion of test particles interacting with a test geodesic radiation flux originating in the equatorial plane of a Schwarzschild space-time and propagating initially in the perpendicular direction. We assume that the interaction with the test particles is modelled by an effective term corresponding to the Thomson-type interaction which governs the Poynting-Robertson effect. After approximating the individual photon trajectories adequately, we solve the continuity equation approximately in order to find a consistent flux density with a certain plausible prescribed equatorial profile. The combined effects of gravity and radiation are illustrated in several typical figures which confirm that the particles are generically strongly influenced by the flux. In particular, they are both collimated and accelerated in the direction perpendicular to the disc, but this acceleration is not enough to explain highly relativistic outflows emanating from some black-hole--disc sources. The model can however be improved in a number of ways before posing further questions which are summarized in concluding remarks.
\end{abstract}



\section{Introduction}

Motion of test particles under the combined effects of gravity and radiation is of obvious astrophysical significance, mainly in the case of the rarified atmosphere around a bright compact source. In the literature, such a motion has mostly been studied while approximating the particle-radiation interaction by a Thomson-like term which specifies, through an effective cross-section constant, what part of the radiation's relative momentum is transferred to the particle. Adopting this approach, we have analyzed the ``Poynting-Robertson effect" of radiation drag in the equatorial plane of the Schwarzschild and Kerr background space-times, for an outgoing or ingoing ``radial" photon flux with zero or non-zero angular momentum \citep{BiniJS-09,BiniGJSS-11}. Then we have also considered \citep{BiniGJS-11} the case of a non-test flux involved in the {\em exact} Vaidya solution, describing a spherically symmetric centre emitting or accreting radiation. These papers may be consulted for a wider review of literature on this topic.

In the meantime, several new contributions to the subject have appeared. \cite{OhKL-10} presented a numerical treatment of particle trajectories in the radiation field of a slowly rotating Kerr-like source, where the existence of equilibrium circular orbits (``suspension" orbits) was confirmed. \cite{Stahl-etal-12} studied the halt and ``levitation" of particles at the corresponding ``Eddington sphere" and discussed its implications for accretion onto a luminous star. In accord with intuition and experience, they concluded that the effective cross section of such a shining source is typically less than the geometric value, because the infall onto the star's surface is prevented by outgoing radiation. In contrast, \cite{OhPK-13} inferred from numerical experiments that luminosity {\em enhances} the effective cross section of a relativistic centre about 4 times. \cite{Stahl-KWA-13} and \cite{MishraK-14} analyzed the response of the matter suspended on the equilibrium ``Eddington sphere" on a sudden luminosity change, mainly aiming at determination of conditions under which ejection from the system may occur.

In the present paper, we consider a radiative flux directed away from the equatorial plane in the ``vertical" direction, in an effort to model the situation which may be generated by a thin accretion disc surrounding a compact gravitational object. We investigate the behavior of test particles above the disc, mainly in the region near the axis of symmetry.
This question has already been tackled several times in the literature in connection with the acceleration/deceleration and axial collimation of astrophysical jets apparently coming out of the above-type accretion systems both on stellar and galactic scales.\footnote
{An up-to-date review of the accretion-disc theory is maintained by \cite{AbramowiczF-13}. The particular issue of jet outflows has been surveyed e.g. by \cite{PudritzHG-12}, with special emphasis put on the supposedly crucial role of magnetic fields.}
In a seminal paper, \cite{BisnovatyiKB-77} calculated the action of radiation on particles in the neighbourhood of a thin disc around a black hole (represented by a Newtonian centre) in their study of various possible consequences of radiation emission on the disc accretion. They mainly analyzed the dependence of particle motion (and of the latter's aftermaths) on the value of luminosity, assuming this is generated by the relativistic version of the Shakura-Shunyaev ``$\alpha$-model" of thin discs due to Novikov and Thorne, and deduced, in particular, that accretion ceases to be possible (at least against the direction of the main energy release) when the luminosity approaches some value around the Eddington one. It was also suggested there that radiation push could ``sow" (weak) electric currents and thus electromagnetic field in plasma due to its stronger effect on electrons than on ions.

Next, \cite{SikoraW-81}, \cite{Piran-82} and \cite{BodoFMT-85} analyzed the radiation acceleration and collimation of test particles or fluid within funnels of thick discs, assuming a Thomson-type interaction. On the other hand, \cite{Phinney-82} argued that ``the greatly enhanced radiation pressure force felt by a relativistic plasma is accompanied by catastrophic Compton cooling and only under extreme conditions can it lead to relativistic bulk velocities." This conclusion was also confirmed by \cite{MeliaK-89} in their study of radiation-drag deceleration of very fast outflows. Then \cite{VokrouhlickyK-91} considered the motion of test particles moving along the symmetry axis of the Schwarzschild or Kerr space-times under the influence of radiation from a thin test disc determined by the Novikov-Thorne model. They propagated the radiation predicted by this model to the location of the particle and there integrated over the latter's local sky, taking into account all the effects of general relativity resulting from the curvature of space induced by a central black hole. The energy-momentum tensor obtained in this manner was then projected onto the particle's four-velocity in order to find the force which the radiation exerts on the particle. The authors concluded that the general relativistic effects on the radiation field (redshift, ray bending, dragging) do not affect the terminal speed of the particle significantly and also did not observe any significant effect of radiation on the axial pre-collimation of particles launched from the surface of the disc. They noticed, however, that the results did depend strongly on the luminosity profile of the disc primarily through the rotation of the central object and pointed out that different conclusions might therefore be reached with different disc models.

Since the black holes supposed in astrophysical sources may be spinning rapidly, the question also appeared naturally whether the rotating (Kerr) space-time geometry could not itself accelerate and/or axially (pre-)collimate outflows emerging from its inner region --- see \cite{BicakSH-93}, \cite{deFeliceZ-00}, \cite{Williams-04}, \cite{TakamiK-09}, \cite{GarielMMS-10} and \cite{deFreitasPachecoGM-12}. However, today the astrophysical jets are believed to be mainly driven by magneto-hydrodynamical effects (e.g. \citealt{PudritzHG-12}).

In the meantime, the interest in {\em radiation} acceleration of jets has continued and more astrophysically quite sophisticated treatments have appeared since then, incorporating radiation from specific models of accretion discs, a more realistic description of the radiation-particle interaction (dependent on energy and taking into account heating of the particle as well as its radiation losses), specific particle content of the outflow (electron-proton or/and electron-positron jets, for example), magnetic fields or/and special geometry of the interaction region (``funnels" of thick accretion discs, in particular) --- see \cite{SikoraSBM-96}, \cite{InoueT-97}, \cite{MadauT-00}, \cite{ChattopadhyayCh-02}, \cite{OriharaF-03}, \cite{FukueA-06}, \cite{TakeuchiOM-10}, \cite{KumarChM-14}, \cite{Cao-14} and their references.
Let us conclude this overview by \cite{KoutsantoniouC-14} who have studied the influence of disc radiation on dynamics of particles at the inner edge, placing the accretion system around a rapidly rotating Kerr black hole. They found that for particles around the innermost stable circular orbit the effect of radiation becomes almost entirely azimuthal and that, interestingly and contrary to a standard intuition, it rather changes from {\em drag} to acceleration. This should enhance the efficiency of a ``cosmic battery" mechanism in which the radiation push might trigger the jet outflows indirectly, through the production of magnetic field.

We would like to compare these various results (especially those of \citealt{VokrouhlickyK-91}) with what can be found using the approach we have taken in previous papers. Restricting to the Schwarzschild case for simplicity now, in Section~\ref{flux} we {\em prescribe} the radiative flux to be emanating perpendicularly from the equatorial plane (where the thin accretion disc is imagined to lie) and study its properties, and then its interaction with the test particles in Section~\ref{interaction}. Then in Section~\ref{approximation} we suitably approximate the photon trajectories, choose the equatorial energy-density profile of the flux and extend it off the equatorial plane by approximating the conservation laws which govern its behavior. Then we add the contributions from two opposing radiation streams which---due to the symmetry---pass through each point and then compute their effect on the particle motion. Numerical examples are given in Section~\ref{examples} and the concluding section ends with several remarks and plans for further study. Note that we use geometrized units in which $c=1$ and $G=1$, Greek indices take the values $0,1,2,3$ and Latin indices $1,2,3$, and partial derivatives are indicated by a comma.

\section{``Vertical" geodesic radiation flux in a Schwarzschild field}
\label{flux}

\begin{figure}
\includegraphics[width=\columnwidth]{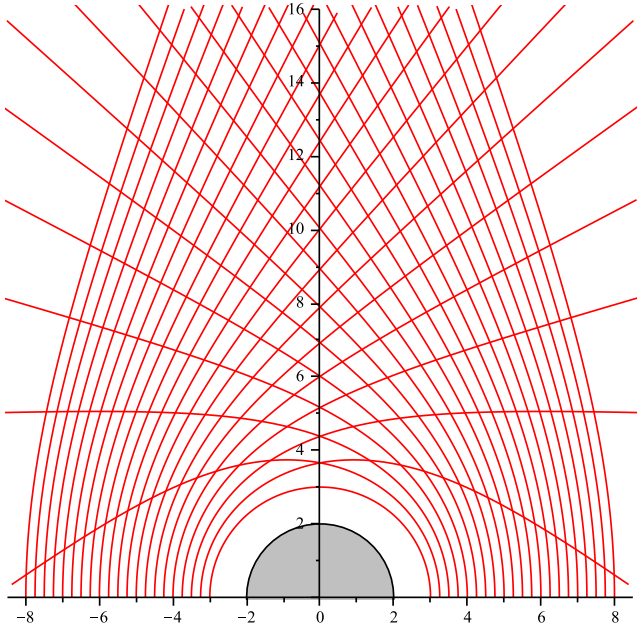}
\caption
{The flux of free test photons emitted perpendicularly from the Schwarzschild equatorial plane into the ``upper" half-space, as plotted in coordinates $(\pm r\sin\theta,r\cos\theta)$ for a typical ``meridional" plane through the symmetry axis. Through each spatial location pass two rays, except for the axis where each point is crossed by photons coming from an entire equatorial circle. Photons starting from below $r=3M$ fall into the black hole (the grey circle) and are not shown. In all the plots the axis values are given in the units of $M$.}
\label{vertical-photons}
\end{figure}

A thin accretion disc in the equatorial plane would certainly emit radiation in all directions, but it is perhaps a natural zero-order approximation to assume that most of the flux is directed perpendicularly to this plane. Actually, \cite{BisnovatyiKB-77} calculated, for a particle at any given location, the radiation force by integrating contributions from all directions over the whole Novikov-Thorne disc, and found a ``cosine law" peaked along the vertical axis. However, since the disc itself orbits around the centre (in fact extremely fast in the case of very compact centre), the radiation it emits should have some angular momentum, but we will still set this angular momentum to zero here, not only for simplicity, but mainly because otherwise the radiation could not reach the vicinity of the axis. Thus we assume that a static axially symmetric thin test disc lies in the equatorial plane $\theta=\pi/2$ of the Schwarzschild black hole outside some radius greater than the horizon radius, and that the photons emanate from it in the perpendicular directions and follow geodesics. Due to the axial symmetry, each spatial point above the disc is then crossed by two rays,\footnote
{In fact by {\em four} of them, if the fluxes starting from both faces of the disc were taken into account. We will only consider one of them, however. Also, we neglect higher-order rays which reach the given point after making one or more full circuits around the black hole.}
except for the symmetry axis $\theta=\pi/2$ where at each point all the rays starting from a certain circular loop meet in a ``caustic", as illustrated in Fig.~\ref{vertical-photons}.

Writing the metric in the standard Schwarzschild form
\[{\rm d}s^2=-N^2{\rm d}t^2+\frac{{\rm d}r^2}{N^2}+r^2({\rm d}\theta^2+\sin^2\theta\,{\rm d}\phi^2)\]
with $N^2=1-\frac{2M}{r}$,
our photons with zero axial angular momentum have non-zero four-momentum components given by
\begin{equation}  \label{photon-momentum}
  p^t=\frac{\sqrt{\cal K}}{N^2 b} \,, \quad
  p^r=\epsilon^r\,\frac{\sqrt{\cal K}}{r}\,\sqrt{\frac{r^2}{b^2}-N^2} \;, \quad
  p^\theta=\epsilon^\theta\,\frac{\sqrt{\cal K}}{r^2} \;,
\end{equation}
where $b\equiv\sqrt{\cal K}/E$ denotes their impact parameter, $E=-p_t$ is their energy at infinity and ${\cal K}=(p_\theta)^2$ is their ``Carter constant", all remaining conserved along the rays; the signs $\epsilon^r\equiv\pm 1$ and $\epsilon^\theta\equiv\pm 1$ fix the orientation of the meridional-motion components. At each location above the circular photon orbit at $r=3M$, this ``null dust" constitutes an outward flux ($\epsilon^r\!=\!+1$) which would admittedly drag any test particle along. Besides the shape of the photon trajectories, the angular distribution of the flux and thus of the particle acceleration/deceleration---in particular, the eventuality that the particles might be driven into a collimated outflow---depend on the ``luminosity profile" fixed in the equatorial plane, namely on a chosen equatorial radial profile of the constants of the motion $E_{\rm eq}=E(r_{\rm eq},\theta\!=\!\pi/2)$ and ${\cal K}_{\rm eq}={\cal K}(r_{\rm eq},\theta\!=\!\pi/2)$ that the photons are endowed with, but mainly on energy density of the flux determined by conservation laws. The constants are constrained by the requirement that the rays depart orthogonally from the equatorial plane, namely by the condition $p^r=0$ there, which takes the form
\begin{equation}  \label{EK-constraint}
  \frac{E_{\rm eq}^2}{{\cal K}_{\rm eq}}\,r^2_{\rm eq}-1+\frac{2M}{r_{\rm eq}}
  \equiv\frac{r_{\rm eq}^2}{b^2}-N^2_{\rm eq}
  =0 \;\; \Rightarrow \;\;
  b=\frac{r_{\rm eq}}{\sqrt{1-\frac{2M}{r_{\rm eq}}}} \;.
\end{equation}

\begin{figure}
\includegraphics[width=\columnwidth]{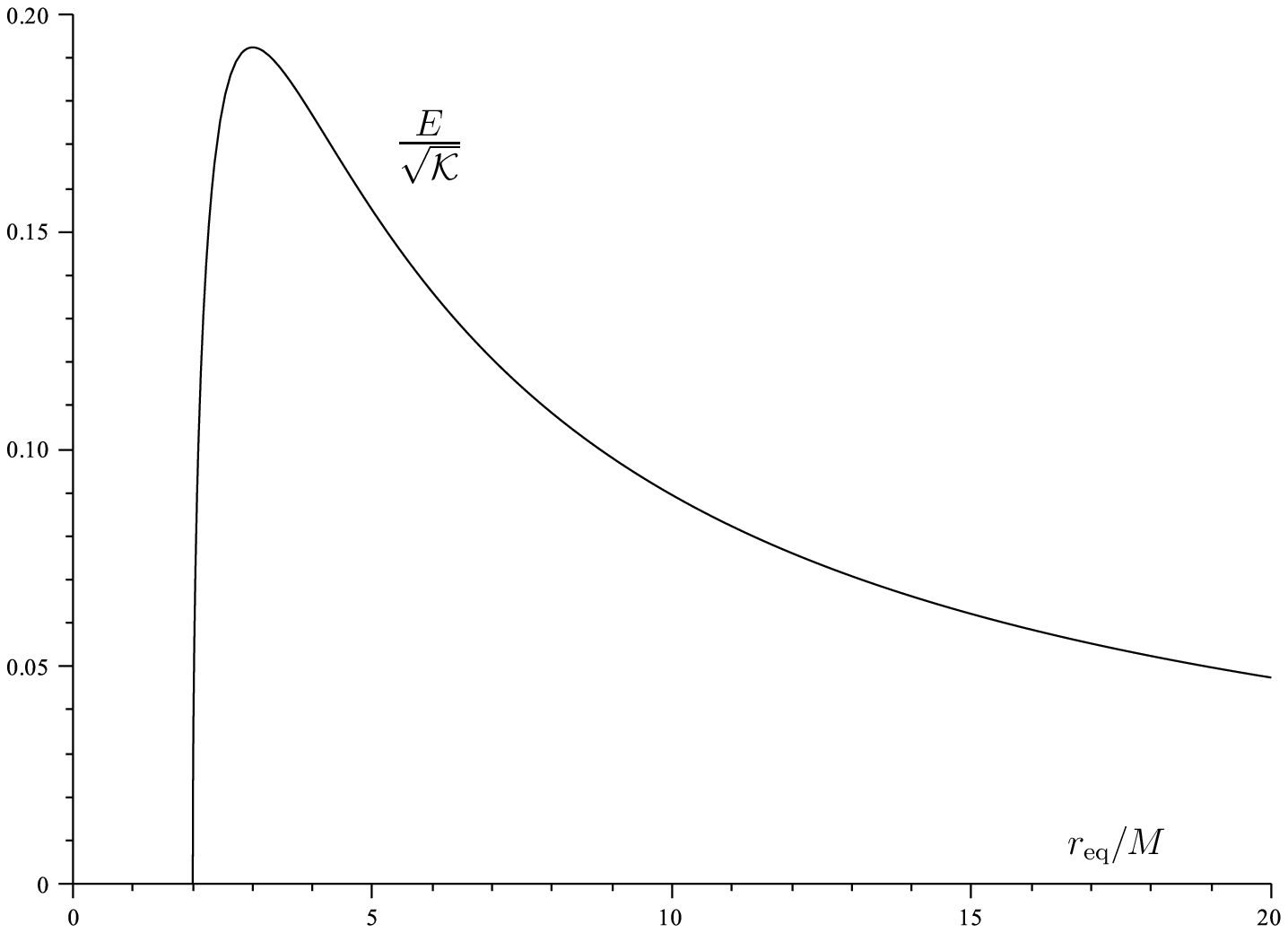}
\caption
{The energy profile $E=-p_t$ (expressed in units of $\sqrt{\cal K}/M$) of photons shot perpendicularly from the equatorial plane as a function of the initial equatorial radius $r_{\rm eq}$, for the case when ${\cal K}$ is chosen to be independent of $r_{\rm eq}$. This profile is just the reciprocal of the well known effective potential for free radial motion of massless particles in the Schwarzschild space-time. It starts from zero at $r_{\rm eq}=2M$, has a global maximum of $(3\sqrt{3}M)^{-1}$ at $r_{\rm eq}=3M$ and then falls back to zero asymptotically.}
\label{E-profile}
\end{figure}

This radius-dependent constraint implies that only one of the two constants of the motion may be chosen to have the same value {\em across} all the rays, thus determining the other as a function of the initial radius. Regarding the supposed accretion-disc temperature profiles, it does not seem wise to endow all the photons with the same energy $E$, and also the resulting profile of ${\cal K}_{\rm eq}(r_{\rm eq})$ implied by the constraint is not very plausible. We will therefore fix the Carter constant ${\cal K}$ instead, which implies that the energy profile has to read $E_{\rm eq}/\sqrt{\cal K}=N_{\rm eq}/r_{\rm eq}$; it is illustrated in Fig.~\ref{E-profile}. Since real accretion discs are supposed to be considerably hotter at smaller radii, but this property is somewhat opposed by larger redshift there with respect to infinity, the profile seen in the figure seems to be a reasonable choice, in particular it properly goes to zero at the horizon. Rough as the ${\cal K}={\rm const}$ choice may seem, the corresponding energy profile in Fig.~\ref{E-profile} actually well resembles the temperature profiles obtained from standard models of thin discs --- see, for example, \cite{BhattacharyyaTB-01} who compared, in their figure 7, the temperature profiles for discs around a Newtonian centre, around a Schwarzschild black hole and around neutron- or strange-star models employing different equations of state; useful plots were also presented by \cite{Perez-etal-13}, showing the temperature profiles of Shakura-Sunyaev and Novikov-Thorne thin discs around a Schwarzschild centre in general relativity and in simple $f(R)$ theories (see their figures 10 and 14, in particular).

Now, taking ratios of Eqs.~(\ref{photon-momentum}) with $\epsilon^r=+1$ leads to
\begin{equation}  \label{dtheta/dr}
  \frac{{\rm d}\theta}{{\rm d}r}
  =\frac{\epsilon^\theta}{r}\,\frac{1}{\sqrt{\frac{E^2}{\cal K}\,r^2-N^2}}
  =\frac{\epsilon^\theta}{\sqrt{r}}\,\frac{1}{\sqrt{{r^3}/{b^2}-r+2M}}
\end{equation}
which can be further expressed in terms of the initial equatorial radius by substituting $E_{\rm eq}=\sqrt{\cal K}\,N_{\rm eq}/r_{\rm eq}$ for $E$ to obtain
\begin{align}
  \frac{{\rm d}\theta}{{\rm d}r}
  &= \frac{\epsilon^\theta\,\frac{r_{\rm eq}}{r}}
          {\sqrt{N_{\rm eq}^2 r^2-N^2 r_{\rm eq}^2}} \nonumber \\
  &=\frac{\epsilon^\theta\,r_{\rm eq}^{3/2}}
         {\sqrt{r(r-r_{\rm eq})}\;
          \sqrt{r(r+r_{\rm eq})(r_{\rm eq}-2M)-2Mr_{\rm eq}^2}}
    \;.  \label{dtheta/dr,req}
\end{align}

In order to integrate this equation, notice the cubic polynomial inside the square root in the denominator of the rightmost side of equation (\ref{dtheta/dr}). Its graph is everywhere concave upward, having two real roots above the horizon, one below and one above $3M$. All our photons start moving perpendicularly from the equatorial plane at $r>3M$ and escape to infinity, which means that the integration is performed just from the outer root $r_{\rm eq}$ (turning point of radial motion) up to a desired radius $r>3M$.
Assuming without loss of generality that the photons start moving ``upwards" from the equatorial plane (i.e., that $\epsilon^\theta\!=\!-1$ initially), one finds the solution formula obtained by Darwin \citep{Darwin-59} (see also equation (260) in Chapter 3 of \citealt{Chandrasekhar})\footnote
{Written in this way, the formula is valid only until the elliptic-integral term reaches $3\pi/2$. Such a value is only reached for photons starting from $r_{\rm eq}<3.09M$, however.}
\begin{align}
  -\epsilon^\theta \theta(r)
  &= \frac{\pi}{2}
     -\frac{2\,\sqrt{r_{\rm eq}}}{[(r_{\rm eq}-2M)(r_{\rm eq}+6M)]^{1/4}}
      \left[K(k)-F(\chi,k)\right] \label{theta(r)} \\
  &= \frac{\pi}{2}
     -\frac{2\,\sqrt{r_{\rm eq}}}{[(r_{\rm eq}-2M)(r_{\rm eq}+6M)]^{1/4}}\;
      F(\chi',k) \,,  \label{theta(r)'}
\end{align}
where $F(\chi,k)=\int_0^\chi\frac{{\rm d}\alpha}{\sqrt{1-k^2\sin^2\alpha}}$
is the elliptic integral of the 1st kind, with amplitude $\chi$ and modulus $k$ given by
\begin{align}
  \sin^2\chi
   &= 1-\frac{1}{k^2}\,
        \frac{2M\left(1-\frac{r_{\rm eq}}{r}\right)}{\sqrt{(r_{\rm eq}-2M)(r_{\rm eq}+6M)}}
    \;, \\
  2k^2
   &= 1-\frac{r_{\rm eq}-6M}{\sqrt{(r_{\rm eq}-2M)(r_{\rm eq}+6M)}} \;,  \label{2k2}
\end{align}
and $K(k)=F(\pi/2,k)$ is its complete version.
One can check immediately that $F(\chi,k)$ only reduces to $K(k)$ at the starting point, where $r=r_{\rm eq}$ and so $\chi=\pi/2$, which correctly yields $\theta(r\!=\!r_{\rm eq})=\pi/2$.
The second expression (\ref{theta(r)'}) contains a different amplitude $\chi'$ which is related to $\chi$ by
\begin{eqnarray}
\lefteqn{
  \sin^2\chi'
   = \frac{1-\sin^2\chi}{1-k^2\sin^2\chi}} \nonumber \\
\lefteqn{\quad
   = \frac{1}{k^2}\,
      \frac{4M\left(1-\frac{r_{\rm eq}}{r}\right)}
           {\sqrt{(r_{\rm eq}-2M)(r_{\rm eq}+6M)}+r_{\rm eq}-2M-4M\frac{r_{\rm eq}}{r}}
  \;.}
\end{eqnarray}
The complementary modulus $k'$ which is related to $k$ by $k'^2=1-k^2$ is given by the same expression (\ref{2k2}) as $k$, just with a {\em plus} sign after the 1; their product is therefore quite short,
\begin{equation}
  k^2 k'^2=\frac{4M\,(r_{\rm eq}-3M)}{(r_{\rm eq}-2M)(r_{\rm eq}+6M)} \;.
\end{equation}

The latitude $\theta$ of all our photons decreases from $\pi/2$ until they cross the symmetry axis $\theta=0$. From there $\theta$ increases back, which is ensured by the sign $\epsilon^\theta$ on the left hand side of (\ref{theta(r)}). In describing the photon trajectories, this sign should only appear in front of $\sin\theta$ terms: actually, one can effectively treat $\sin\theta$ (as well as $\theta$ itself) as positive for photons which have not yet crossed the axis while as negative for those which have already crossed it. Such a distinction will be important in evaluation of the photon effect on the particle, because at each (non-axial) point the particle interacts with just {\em two} photons --- one approaching the axis and one already receding from it (the latter started from smaller equatorial radius than the former, so it has been bent more).

Unfortunately, equation (\ref{theta(r)}) represents only an {\em implicit} relation between $(r,\theta)$ and $r_{\rm eq}$ and can only be solved numerically for $r_{\rm eq}$ in general.
More precisely it determines the trajectory of the photon as parametrized by its starting radius $r_{\rm eq}$, which can in principle be inverted to ``reconstruct" $r_{\rm eq}$ as a function of the actual photon's position $(r,\theta)$ (this inversion is unique, at least if restricting to $r_{\rm eq}$ larger than a certain radius slightly above $3M$ in order to discard photons which make more than one full revolution in $\theta$ before reaching infinity). Being able to trace $r_{\rm eq}$ from the actual position $(r,\theta)$ within the photon flux, one then also learns the distribution of photon energy $E$ in space (and thus of their impact parameter $b\!=\!\sqrt{\cal K}/E$ as well), because $r_{\rm eq}$ is uniquely related to $E_{\rm eq}$ (at least at $r_{\rm eq}\!>\!3M$, which is relevant), namely
\[E(r,\theta)=E_{\rm eq}{}_{(r,\theta)}=\sqrt{\cal K}\,(N_{\rm eq}/r_{\rm eq})_{(r,\theta)}\]
where ${\cal K}$ is an absolute constant, and the subscript notation indicates this implicit relationship.

\subsection{Energy-momentum tensor}

The radiation flux will be described as an incoherent ``null dust" with energy-momentum tensor
\begin{equation}  \label{Tmunu}
  T^{\mu\nu}=\Phi^2 p^\mu p^\nu \,,
\end{equation}
where $\Phi^2(r,\theta)$ scales the radiation energy density. The latter has to be fixed by the conservation law ${T^{\mu\nu}}_{;\nu}=0$ after choosing a certain profile on some surface stretching across the rays; in our case, it is natural to choose the equatorial profile $\Phi^2_{\rm eq}=\Phi^2(r_{\rm eq},\theta\!=\!\pi/2)$.
For an incoherent radiation flux, this implies
\[0=(\Phi^2 p^\mu p^\nu)_{;\nu}
   =\Phi^2 {p^\mu}_{;\nu}p^\nu+p^\mu(\Phi^2 p^\nu)_{;\nu}
   =p^\mu(\Phi^2 p^\nu)_{;\nu} \;,\]
because the photon congruence is geodesic: ${p^\mu}_{;\nu}p^\nu=0$.
Hence for the particular ``vertical" flux chosen in the previous section, ${T^{\phi\nu}}_{;\nu}=0$ is satisfied trivially due to $p^\phi=0$ (in the equatorial plane, ${T^{r\nu}}_{;\nu}=0$ also holds automatically, because $p^r=0$ there), while the other components reduce to a single common condition
\begin{equation}  \label{Div(Phi2.p)=0}
  (\Phi^2 p^\nu)_{;\nu}=0
  \quad \Longrightarrow \quad
  (\Phi^2)_{,\nu}p^\nu=-\Phi^2 {p^\nu}_{;\nu}
\end{equation}
which says that the evolution of $\Phi^2$ {\em along the photon congruence} is tied to the latter's expansion ${p^\nu}_{;\nu}$; somewhat more explicitly,
\begin{equation}
  (\Phi^2)_{,r}\,p^r+(\Phi^2)_{,\theta}\,p^\theta
  =-\frac{\Phi^2}{\sqrt{-g}}
    \left[(\sqrt{-g}\,p^r)_{,r}+(\sqrt{-g}\,p^\theta)_{,\theta}\right].
\end{equation}
An even more explicit equation follows using equation (\ref{photon-momentum}) to substitute for $p^i$. In doing so, one has to realize that the differentiation is performed in a {\em general} direction, not just along the photon rays, so it must be performed with every quantity that is not constant all over the radiation field; in particular, this even applies to the constants of geodesic motion (which are only constant {\em along the rays}) unless they are same for {\em all} the rays. With our choice made in previous section, this means that one has to consider the energy $E$ to be a function of $(r,\theta)$, whereas ${\cal K}$ is left constant since it has been chosen to be the same for all photons. The conservation condition can thus be expressed as
\begin{equation}  \label{Div(Phi2.p)=0,partial}
  (r^2\Phi^2 p^r)_{,r}\sin\theta+\epsilon^\theta\sqrt{\cal K}\,(\Phi^2\sin\theta)_{,\theta}=0
\end{equation}
or after substitution for $p^r=({\sqrt{\cal K}}/{r})\,\sqrt{{r^2}/{b^2}-N^2}\,$
\begin{equation}
  \left(r\,\Phi^2\,\sqrt{\frac{r^2}{b^2}-1+\frac{2M}{r}}\;\sin\theta\right)_{\!\!,r}+
  \epsilon^\theta(\Phi^2\sin\theta)_{,\theta}=0 \;.
\end{equation}
Expanding the product derivative and dividing through, one finds
\begin{equation}  \label{equation-for-Phi}
  \left(r\,\Phi^2\right)_{,r}
  +\frac{\epsilon^\theta\,(\Phi^2\sin\theta)_{,\theta}}{\sqrt{\frac{r^2}{b^2}-1+\frac{2M}{r}}\;\sin\theta}
  +\Phi^2\,\frac{\frac{r^2}{b^2}-\frac{r^3}{b^3}\,b_{,r}-\frac{M}{r}}
                {\frac{r^2}{b^2}-1+\frac{2M}{r}}
  =0 \,,
\end{equation}
where $b=r_{\rm eq}(1-2M/r_{\rm eq})^{-1/2}$ according to the constraint (\ref{EK-constraint}).
Apparently it is correct to keep the $\epsilon^\theta$ sign in the equation in order to distinguish between the flux approaching the axis and its successor continuing after crossing the axis, since otherwise the $\theta$-derivative would jump across the axis due to the reversal of the $\partial x^\mu/\partial\theta$ orientation.

\section{Interaction of a test particle with the radiation flux}
\label{interaction}

The aim of this paper is to check whether the radiation flux from the disc could not accelerate test particles and/or collimate them in the direction perpendicular to the disc plane. Hence, consider a test particle moving in the Schwarzschild field and influenced by interaction with radiation described by the null dust energy-momentum tensor of equation (\ref{Tmunu}). If the interaction is dominated by Thomson scattering, it is convenient to approximate its effect on the particle in terms of the fraction of transferred radiation momentum, as seen in the particle's rest frame, i.e., by the equation of motion
\begin{equation}  \label{equation-of-motion}
  \frac{{\rm D}u^\mu}{{\rm d}\tau}
        =-\tilde{\sigma}\,(\delta^\mu_\nu+u^\mu u_\nu)\,T^\nu{}_\lambda u^\lambda
        =\tilde{\sigma}\,\Phi^2\hat{E}\,\hat{p}^\mu \,,
\end{equation}
where $\tau$, $u^\mu$ and $a^\mu={{\rm D}u^\mu}/{{\rm d}\tau}$ are the test-particle proper time, four-velocity and four-acceleration, $(\delta^\mu_\nu+u^\mu u_\nu)$ is the projector onto the particle's local rest space and $\tilde{\sigma}$ is an effective constant scaling the interaction strength (with dimensions of length); the second version of the right hand side is written in terms of the relative photon energy and momentum with respect to the particle, $\hat{E}\equiv -p_\nu u^\nu$ and $\hat{p}^\mu$, which result from the decomposition
\begin{equation}
   p^\mu=-u^\mu u_\nu p^\nu+(\delta^\mu_\nu+u^\mu u_\nu)\,p^\nu
        =\hat{E}u^\mu+\hat{p}^\mu \,.
\end{equation}

Despite the elegant form (\ref{equation-of-motion}) of the equation of motion, the expressions for $\hat{E}$ and $\hat{p}^\mu$ make it rather cumbersome to be given explicitly here, the only simplification occurring thanks to the zero azimuthal motion of photons, $p_\phi=0$.
Note, however, that although both the gravitational and radiation fields are axially symmetric ($T_{\mu\phi}=0$), the force does contain a nonzero azimuthal component if the particle's velocity has some, because of the projection term $u^\phi\,T_{\nu\lambda}u^\nu u^\lambda$.

\begin{figure*}
\includegraphics[width=\textwidth]{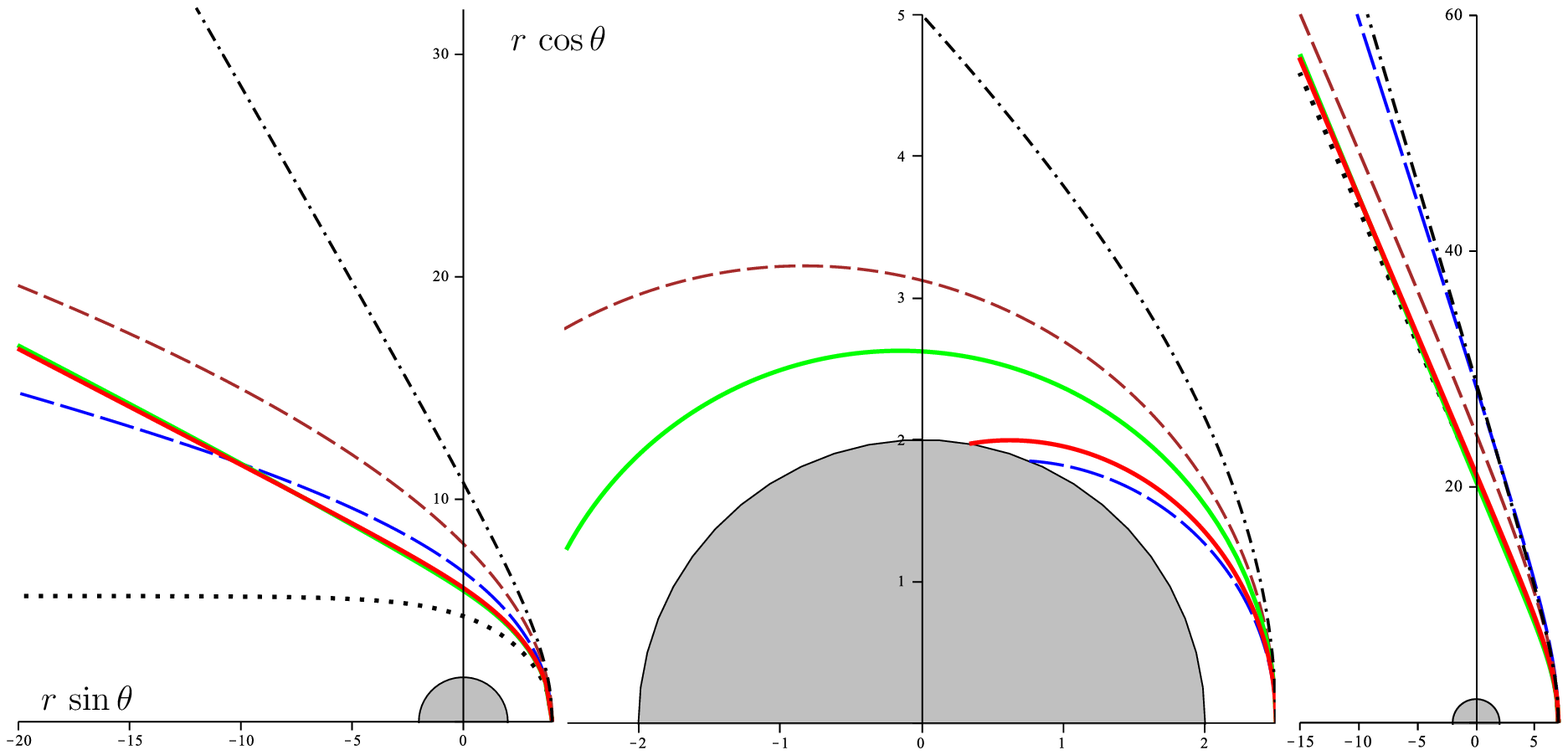}
\caption
{Meridional-plane plots of 3 photon trajectories, as represented by the exact formula (thick red) and by several approximations we have considered: Beloborodov's approximation is black dotted (it is only applicable above $4M$); approximation by a suitably adjusted hyperbola is brown dashed; pseudo-Newtonian result using the potential $V\!=\!-\frac{M}{r}\left(1+\frac{3M}{r}\right)$ is long-dashed blue; our approximation given by formula (\ref{our-approximation}) and used below is thick green; and the result following from standard linearization in $M$ of the exact formula (the worst approximation) is dash-dotted. The photon starts from $r_{\rm eq}=4M$ (left), $r_{\rm eq}=7M$ (middle) and $r_{\rm eq}=2.5M$ (right). Our approximation almost coincides with the exact trajectories for photons starting down to some $3.8M$ and even below $3M$ (down to the very horizon) it does not yield nonsense (it follows a constant radius there approximately).}
\label{approximations}
\end{figure*}

\section{Approximating the photon trajectories}
\label{approximation}

In order to evaluate the effect of the photon flux on the test particle at a given point $(r,\theta)$, one would have to solve equation (\ref{theta(r)}) for $r_{\rm eq}$ and find the photon energy $E=\sqrt{K}/b$ or the impact parameter $b$ (and thus the momentum of the incoming photon) there. Then one would have to determine $\Phi^2(r,\theta)$ at that point by solving the continuity equation (\ref{Div(Phi2.p)=0}) with a prescribed ``velocity" $p^\mu(r,\theta)$. If we do not want to resort to pure numerics to accomplish these two steps, we can consider trying analytic approximations.

A natural approach is to linearize the problem in some small parameter. We have $b>r_{\rm eq}$ and, restricting to the astrophysically relevant case $r_{\rm eq}>3M$, also $r\geq r_0>3M$, while $r$ and $b$ are less clearly related: all the photons start from $r<b$, but quickly get to $r>b$ and then even to $r\gg b$. One may linearize consistently in several small parameters, but the most frequent is the linearization in $M$ which is always the smallest one.
However, Darwin's exact solution (\ref{theta(r)}) is often better approximated by an ad hoc formula rather than by applying some general approximation scheme; this is mainly true at low radii where the weak-field linearizations give too ``weak" result (we will discuss this issue in more detail elsewhere \citep{Semerak-15}). A good example is the formula provided by Beloborodov \citep{Beloborodov-02} which is often used in the accretion-disc community. Another usable possibility is to approximate the photon meridional-plane trajectory by a suitably adjusted hyperbola. Choosing correctly the asymptotic angle $\theta_\infty$ along which the photon approaches radial infinity, such a hyperbola may be the best approximation at large distances, though close to a horizon it is again bent less than the actual relativistic trajectory. Given the symmetry of our radiation field, both these options can only be used for photons that make less than a $90^\circ$ change in direction, however. Still another possibility is to use a pseudo-Newtonian approach and simulate the Schwarzschild field by a suitably modified Newtonian-type potential. Various forms of such a potential have been suggested, starting from the well known cases of Paczy\'nski and Wiita, $V\!=\!-\frac{M}{r-2M}\,$ (also used in some of the papers cited in the Introduction), or Nowak and Wagoner, $V\!=\!-\frac{M}{r}\left(1-\frac{3M}{r}+\frac{12M^2}{r^2}\right)$; see e.g., the form $V\!=\!-\frac{M}{r}\left(1+\frac{\alpha M}{r}\right)$, with constant $\alpha$, advocated by \cite{Wegg-12} recently (specifically with $\alpha\!=\!3$) which has also proven quite satisfactory in our photon-motion problem.

Actually, it is possible to design a number of rather accurate approximations of the photon trajectories. However, the photon motion is not the full story here: we also have to employ its description (inverted for $r_{\rm eq}$, and thus for $b$) in the continuity equation (\ref{equation-for-Phi}) and then solve the latter for the flux density $\Phi^2$. Although a chosen approximation may allow for a tractable inversion, this often makes the continuity equation too difficult to solve, even after linearization in $M$.

One approximation which leads to a solvable form of the continuity equation is reached by the usual linearization in $M$. \footnote
{Real accretion discs certainly radiate in a much less symmetric and regular way than we consider here, so although it is always nice to have a self-consistent and ``exact" solution, in this case it is clearly sufficient to use any reasonable approximation, at least when trying to determine the flux density from conservation laws. Special attention is only required in the region close to the horizon where approximations may misrepresent the picture heavily or even lead to errors when applied within the {\em exact} background field.}
Linearizing thus Darwin's formula (\ref{theta(r)}) gives
\begin{equation}  \label{sin(theta)=lin}
  -\epsilon^\theta\sin\theta=
  \frac{r_{\rm eq}}{r}-\frac{M}{r}\,\frac{(r-r_{\rm eq})(2r+r_{\rm eq})}{r_{\rm eq}\,r}
  +O(M^2)
\end{equation}
which inverts to
\begin{equation}  \label{req,no-linearization}
  \frac{2r_{\rm eq}}{r}=
  \frac{-\epsilon^\theta r\sin\theta\!-\!M\!+\!\sqrt{(\epsilon^\theta r\sin\theta\!+\!M)^2\!+\!8M(r\!+\!M)}}{r+M} \,.
\end{equation}
Note that on the axis the latter yields
\[r_{\rm eq}(\theta\!=\!0)=\frac{4Mr}{M\!+\!\sqrt{M(8r\!+\!9M)}}
                          =\sqrt{2Mr}-\frac{M}{2}+O(M^{3/2}) \,.\]
These equatorial radii $r_{\rm eq}(\epsilon^\theta=-1)$, $r_{\rm eq}(\epsilon^\theta=+1)$ should then be substituted into the impact parameter $b=\!r_{\rm eq}/N_{\rm eq}$ and this in turn into Equation (\ref{photon-momentum}) in order to find momenta of the two photons which hit the particle at the given location $(r,\theta)$.

A comparison of several approximations is presented in Fig.~\ref{approximations}. Meridional plots of three photon trajectories are shown there, as represented 1) by the exact formula (thick red curve), 2) by Beloborodov's approximation, applicable above $4M$ (black dotted; it is included mainly as a benchmark), 3) by a suitably adjusted hyperbola (brown dashed), 4) by a pseudo-Newtonian result using the potential $V\!=\!-\frac{M}{r}\left(1+\frac{3M}{r}\right)$ (long-dashed blue), 5) by the result following from the above linearization in $M$ of the differential equation (dash-dotted; it is the worst approximation), and 6) by the formula
\begin{equation}  \label{our-approximation}
  -\epsilon^\theta
  \sin\theta=\frac{r_{\rm eq}}{r}-
             \frac{M}{r_{\rm eq}-\alpha M}\,
             \frac{(r-r_{\rm eq})(2r+r_{\rm eq})}{(r-\omega M)^2}
\end{equation}
(where $\alpha$ and $\omega$ are real constants)
which we newly suggest and will specifically use with $\alpha=1.77$ and $\omega=1.45$ (thick green).\footnote
{The issue of satisfactory approximation will be more discussed in \cite{Semerak-15}.}
The hyperbola (brown dashed) is chosen to have the same asymptotic latitude $\theta_\infty$ as the orbit provided by Beloborodov's formula (dotted), namely given by $\sin\theta_\infty\!=\!\frac{2M}{r_{\rm eq}-2M}$. In the {\it top plot} of the figure, the photon starts from $r_{\rm eq}=4M$, where it is already quite hard to mimic the exact result by any low-order formula; however, our approximation even there practically coincides with the exact curve. In the {\it bottom left plot}, the photon starts from $r_{\rm eq}=7M$; Beloborodov's and our approximation are almost indistinguishable from the exact ray. At larger radii the approximations gradually coalesce with the exact curve and nothing interesting happens (only the pseudo-Newtonian result gets worse), so we do not show any photon starting from the more remote, weak-field region. In the {\it bottom right plot}, the approximate formulas are subjected to a very tough situation of a photon starting from $r_{\rm eq}=2.5M$. Surprisingly enough, none of them yields a totally unacceptable result (apart from the linearization in $M$ and from Beloborodov's formula which is, however, not applicable below $4M$, so it is not present), the pseudo-Newtonian (blue) curve is even very close to the exact one and shares its black-hole destiny. Better approximations can be found, but typically they must be of higher order, so usually not invertible for $r_{\rm eq}$ and leading to a rather difficult continuity equation.

\begin{figure}
\includegraphics[width=\columnwidth]{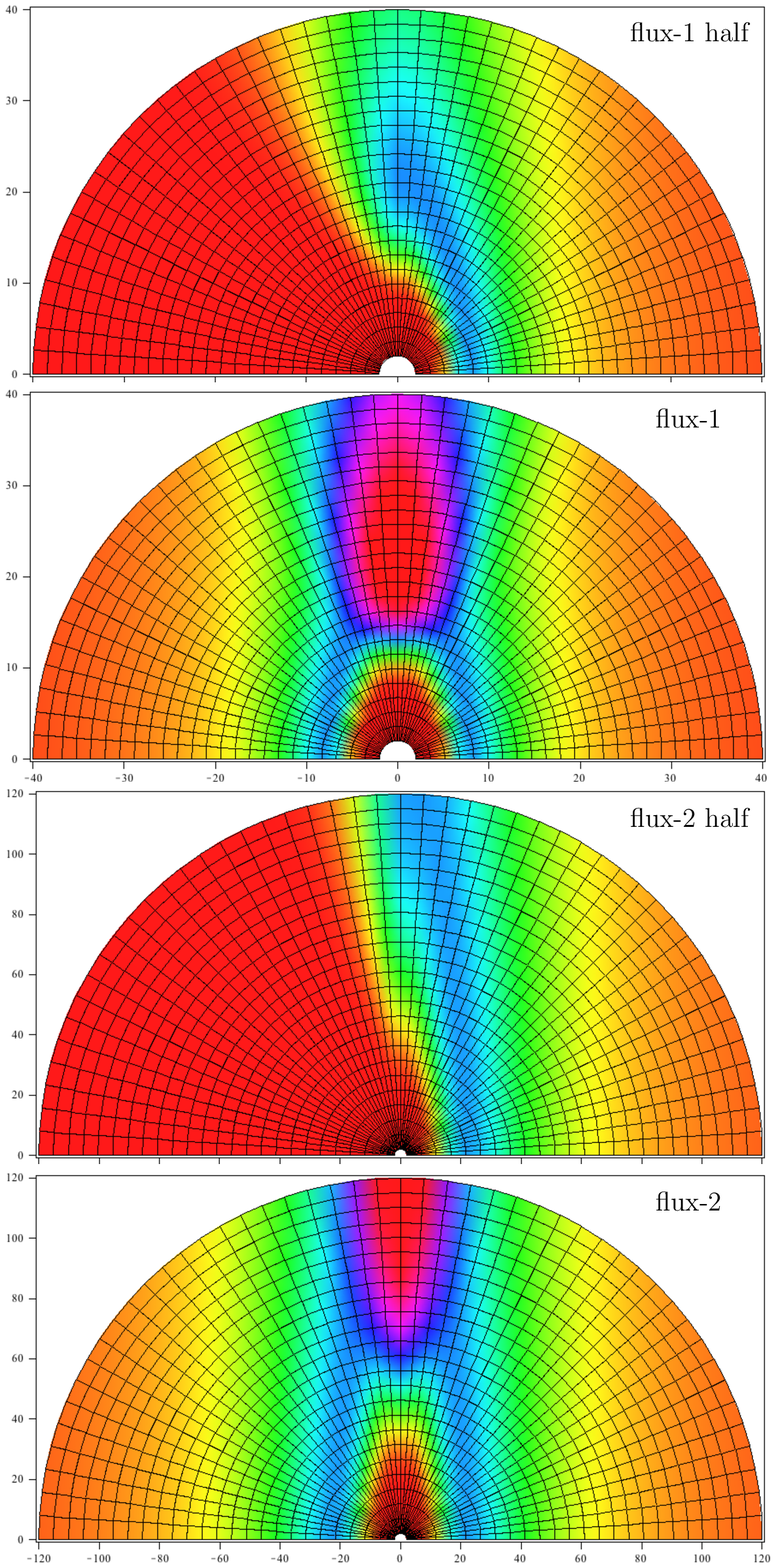}
\caption
{Meridional-plane $(r\sin\theta,r\cos\theta)$ plots of two different solutions for the photon-flux density $\Phi^2(r,\theta)$: the more concentrated photon flux given by $\kappa=32$, $\lambda=2$ ({\it top two plots}) and of the less concentrated photon flux given by $\kappa=64$, $\lambda=1$ ({\it bottom two plots}).
Within both pairs, the first plots show densities of the ``right" halves of the fluxes, namely of those starting from the right-hand half of equatorial plane in the figure: the converging (before-the-axis) part $\Phi^2_-$ (\ref{Phi2-minus}) is plotted in the right quadrant and the diverging (after-the-axis) part $\Phi^2_+$ (\ref{Phi2-plus},\ref{Fplus},\ref{X0Y}) in the left quadrant; both match along $\theta=0$.
The second plots of the pairs show total energy densities $\Phi^2(r,\theta)$, given by superposition $\Phi^2_-+\Phi^2_+$ in all the meridional plane.
Notice the different ranges of the plot pairs. Note also that although the colour shading of both pairs of plots is normalized from red (zero density) across the {\sc HUE} range to violet-red (largest density), the first-flux maximum is about 0.6, while the second-flux maximum is only about 0.046 (see Fig.~\ref{Phi2-plot}).
Horizontal axis corresponds to the equatorial plane ($\theta=\pi/2$), vertical one to the symmetry axis ($\theta=0$); black hole is down in the middle (small light half-circle).}
\label{fluxes-density}
\end{figure}

\begin{figure}
\includegraphics[width=\columnwidth]{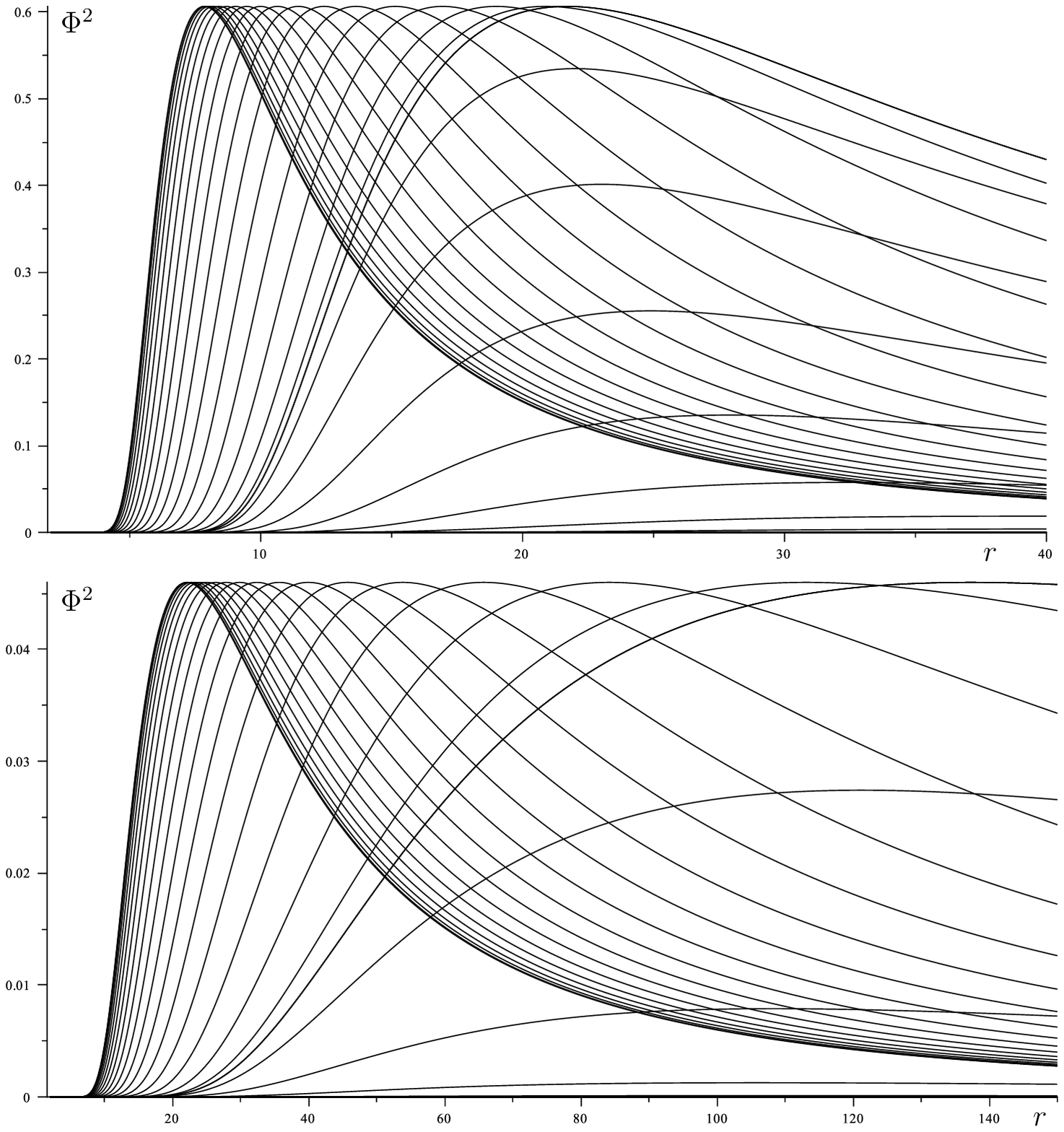}
\caption
{Radial profiles of the two solutions for the photon-flux density shown in Fig.~\ref{fluxes-density}, evaluated along different latitudes. (In other words, radial sections through the 1st and 3rd plots of Fig.~\ref{fluxes-density}, respectively, taken at latitudes scanning the half-circle by $5^\circ$ in a counter-clockwise sense.)
The curves with the same maximal values represent the ``converging" phase of the photon flux $\Phi_-^2(r)$, as given by (\ref{Phi2-minus}), evaluated, from left to right, at latitudes going from the equatorial plane to the symmetry axis ($\theta=90^\circ$, $85^\circ$, $80^\circ$, \dots $0$). In the {\it top} plot, $\kappa=32$ and $\lambda=2$ are chosen; the equatorial density peaks nearby above the innermost stable circular orbit, having very steep falloff towards the horizon and more slow (yet also monotonous) fall towards infinity. The {\it bottom} plot shows a more stretched emission pattern (note the different axes ranges), given by $\kappa=64$ and $\lambda=1$. The photons starting from any given $\phi$ deviate from each other and are outgoing, so it is intuitive that the density profile gets wider when the flux travels from the equatorial plane towards the axis (recall the photon-motion pattern in Fig.~\ref{vertical-photons}). In the right part of the plots, the density of the photon-flux ``diverging" phase $\Phi_+^2(r)$ is shown, as given by (\ref{Phi2-plus},\ref{Fplus},\ref{X0Y}) and matched to $\Phi_-^2(r)$ along the axis; it is evaluated at latitudes going from the symmetry axis back to the equatorial plane by $5^\circ$ again. Only several first profiles of this secondary flux are visible, since the density dissipates quickly when the flux propagates ``behind" the axis, so the curves quickly drop to zero with increasing latitude there.
The radius (horizontal axis) is in the units of $M$, while the flux density is in the units of $M^{-4}$.}
\label{Phi2-plot}
\end{figure}

\subsection{The corresponding photon-flux density}

Using the result (\ref{req,no-linearization}) of the linearization in $M$, one finds that the continuity equation (\ref{equation-for-Phi}) assumes the following two forms
\begin{eqnarray}
\lefteqn{
  \epsilon^\theta=-1 \quad ({\rm ``primary"~flux}, \; \pi/2\geq\theta\geq 0)\,:} \nonumber \\
\lefteqn{\qquad
  r(\Phi^2)_{,r}
  -\left(1+\frac{2+\sin^2\theta}{\sin^2\theta}\,\frac{M}{r}\right)
   (\Phi^2)_{,\theta}\,\tan\theta
  =0 \;,}  \label{continuity,lin-M,-1} \\
\lefteqn{
  \epsilon^\theta=+1 \quad ({\rm ``secondary"~flux}, \; 0\leq\theta<\theta_\infty)\,:} \nonumber \\
\lefteqn{\qquad
  (r^2\Phi^2)_{,r}
  -\frac{(2+\sin\theta)(2-3\sin\theta)}{2\sin^2\theta\,(1-\sin\theta)}\,M\,\Phi^2+} \nonumber \\
\lefteqn{\qquad\qquad\qquad
  {}+2M\,\frac{(\Phi^2\sin\theta)_{,\theta}}{\sin^2\theta\,\sqrt{1-\sin\theta}}
  =0 \;.}  \label{continuity,lin-M,+1}
\end{eqnarray}
Their respective solutions must then be matched on the axis.
It is worth noting that linearizing the continuity equation in $M$, it is much less sensitive to the particular approximation used to describe the rays. For example, for $\epsilon^\theta=-1$, not only the approximation represented by the formula (\ref{sin(theta)=lin}), but all approximations in the above family (\ref{our-approximation})
(and maybe others) lead to the very same linearized continuity equation (\ref{continuity,lin-M,-1}). Note that there are some very good options within the above class of such choices, among them the one given by $\alpha=1.77$ and $\omega=1.45$ which we will use below. Note also that the above is of course true for the $\epsilon^\theta=+1$ case as well, but the classes of ray approximations leading to the same linearized form of the continuity equation are different; in particular, the class just mentioned, when used with $(-\sin\theta)$, does {\em not} yield Equation (\ref{continuity,lin-M,+1}).

The continuity equation (\ref{continuity,lin-M,-1}) for $\Phi^2$ which describes the flux in the $\theta=\pi/2\rightarrow 0$ quadrant is solved by any function $\Phi^2_-(X)$, where
\[
X=\frac{r^2\sin^2\theta+4Mr}{8M^2\exp(2M/r)}-{\rm Ei}\!\left(1,\frac{2M}{r}\right) \,,
\]
and
\[{\rm Ei}\!\left(1,\frac{2M}{r}\right)
  \equiv\Gamma\!\left(0,\frac{2M}{r}\right)
  \equiv\int\limits_{2M/r}^{\infty}\frac{{\rm d}x}{xe^x}\]
is the exponential integral (related with the incomplete $\Gamma$ function).
On the axis it becomes
\begin{equation}
  \Phi^2_-=\Phi^2_-\!\left(\frac{r}{2M}\,\exp\!\left(\!-\frac{2M}{r}\right)
                           -{\rm Ei}\!\left(1,\frac{2M}{r}\right)\right).
\end{equation}
Let us recall our astrophysical motivation, involving radiation from an equatorial accretion disc: (i) real thin accretion discs are assumed to reach close to the innermost stable circular orbit around the compact centre; in the Schwarzschild field this orbit lies at $r=6M$; (ii) the disc temperature is the highest in the region close to its inner edge, so in the disc plane (the equatorial one) the radiation flux peaks somewhere near above $6M$ while falling to zero very quickly (exponentially) towards the horizon and more slowly (probably as $1/r^2$) towards infinity.
Therefore, we can for example choose
\begin{align}
  \Phi^2_- &= \frac{8}{M^4 X}\,\exp\!\left(\!-\frac{\kappa}{X^\lambda}\right), \label{Phi2-minus} \\
      X    &= \frac{r^2\sin^2\theta+4Mr}{8M^2\exp(2M/r)}-{\rm Ei}\!\left(1,\frac{2M}{r}\right),
\end{align}
where $\kappa$ and $\lambda$ are some positive numbers; generically, smaller $\kappa$ and larger $\lambda$ make the profile have a sharper maximum closer to the centre. We will specifically choose ($\kappa=32$, $\lambda=2$) and ($\kappa=64$, $\lambda=1$) for numerical examples; the first case should approximate an accretion disc concentrated towards the innermost stable circular orbit, while the second case corresponds to a disc spread out to larger radii.
At large radii $\Phi^2_-$ falls off as $64/(Mr\sin\theta)^2+O(1/r^3)$, while along the $\theta=0$ axis only as $16/(M^3 r)+O(1/r^2)$.
The above flux profiles really well follow the curves occurring in the accretion-disc literature, see for example figures 9 and 11 in \cite{Perez-etal-13}.

The continuity equation (\ref{continuity,lin-M,+1}) which describes the flux after it has crossed the symmetry axis ($\theta=0\rightarrow\theta_\infty$) is solved by
\begin{align}
  \Phi_+^2 &=
  \frac{F_+(Y)}{M^2 r^2 \; \exp\!\left(\frac{3}{2}\right)}\;
  \frac{\exp\!\left(\frac{3}{2}\,\sqrt{1+\sin\theta}\right)}
       {(1+\sqrt{1+\sin\theta})^2} \;\times \nonumber \\
           & \quad \times
  \left(\frac{(\sqrt{2}+1)\sqrt{1-\sin\theta}}{\sqrt{2}+\sqrt{1+\sin\theta}}\right)^{\!\!\frac{3\sqrt{2}}{4}}
  \!\!,  \label{Phi2-plus}
\end{align}
where $F_+(Y)$ is an arbitrary dimensionless function of
\[
  Y=(2-\sin\theta)\sqrt{1+\sin\theta}-\frac{3M}{r}\;.
\]
On the axis the solution reduces to
\begin{equation}
  \Phi_+^2=\frac{1}{4M^2 r^2}\;F_+\!\left(2-\frac{3M}{r}\right).
\end{equation}

The converging and diverging phases of the flux match together on the axis if $\Phi_+^2=\Phi_-^2$ there, hence if
\[F_+\!\left(2-\frac{3M}{r}\right)
  =\frac{32\,r^2}{M^2 X_{\theta=0}}\,\exp\left(-\frac{\kappa}{X^\lambda_{\theta=0}}\right).\]
One can write this functional relation as
\[F_+(Y_{\theta=0})=\frac{288}{(2-Y_{\theta=0})^2 X_{\theta=0}}\,
                       \exp\left(-\frac{\kappa}{X^\lambda_{\theta=0}}\right),\]
where
\begin{eqnarray*}
\lefteqn{
  X_{\theta=0}\equiv X_{\theta=0}(Y_{\theta=0})
  =\frac{1}{\frac{2M}{r}\,\exp\left(\frac{2M}{r}\right)}-{\rm Ei}\!\left(1,\frac{2M}{r}\right)} \\
\lefteqn{\qquad
  {\rm with} \;\;\; \frac{2M}{r}=\frac{2}{3}\,(2-Y_{\theta=0}) \;.}
\end{eqnarray*}
For a general value of $\theta$, we thus have
\begin{align}
  F_+(Y) &= \frac{288}{(2-Y)^2 X_{\theta=0}(Y)}\,\exp\left(-\frac{\kappa}{X^\lambda_{\theta=0}(Y)}\right),
  \label{Fplus} \\
  X_{\theta=0}(Y) &\equiv \frac{1}{\frac{2}{3}(2-Y)\,\exp\left[\frac{2}{3}(2-Y)\right]}
                          -{\rm Ei}\!\left[1,\frac{2}{3}(2-Y)\right].
  \label{X0Y}
\end{align}
The ``secondary" flux density $\Phi_+^2$ vanishes on the equatorial plane, and at radial infinity it generally falls off as $1/r^2$ (but along the axis only as $1/r$, as known from its matching to $\Phi_-^2$ there).
Both components of the flux are everywhere positive and smooth, having one (global) maximum somewhere between the horizon and infinity.

Sections of the energy-density profile of both parts of the flux, drawn at various latitudes for the two specific cases (one rather concentrated and the other more spread out), are given in Fig.~\ref{Phi2-plot}. Fig.~\ref{fluxes-density} presents the meridional plane energy-density distributions of both the flux solutions; the top plots show half of the fluxes starting from the ``right-hand" half of the equatorial plane, while the bottom plots show total density given by superposition of the top plots with its counter-parts obtained by reflection with respect to the vertical axis. It is clearly seen that the first radiation flux is more concentrated (towards smaller radii). It can be estimated that with a more accurate solution of the continuity equation (namely higher-order in $M$), the flux would more bend around the black hole (consider that the linearization in $M$ makes the centre's field weaker), so it will spread to the axis slightly sooner (i.e., at smaller radii) than in the present plots and the secondary flux component (after crossing the axis) would be correspondingly stronger.

\begin{figure*}
\includegraphics[width=\textwidth]{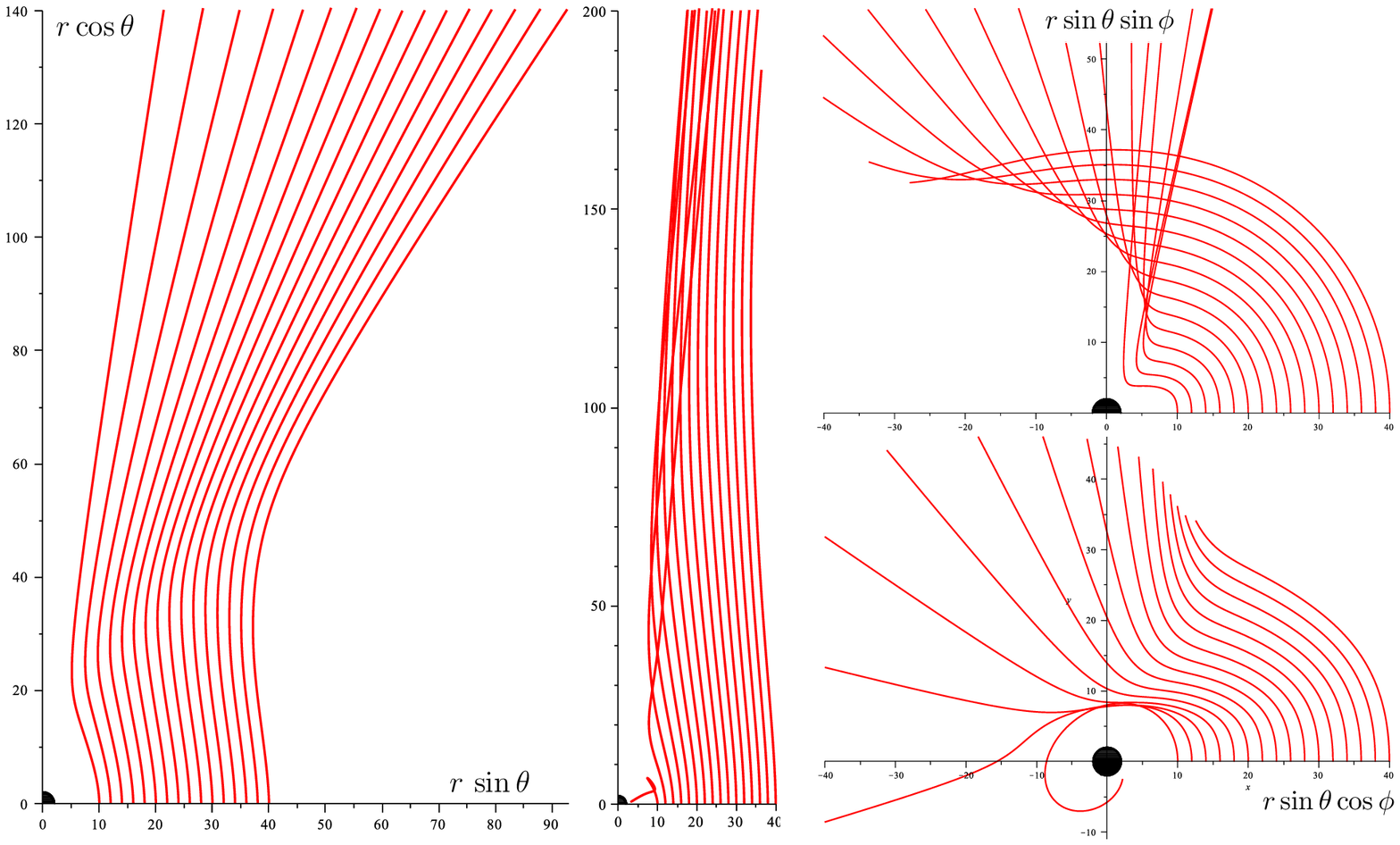}
\caption
{Effect of radiation (and gravity) on particles released at $r=40M$, $38M$, $36M$, \dots, $10M$ from the equatorial plane, with only the azimuthal component of initial velocity non-zero and given by the Keplerian value. The meridional plane $(r\sin\theta,r\cos\theta)$ projections of the particle motion are shown for the more concentrated disc ($\kappa=32$, $\lambda=2$) in the left plot and for the more spread-out disc ($\kappa=64$, $\lambda=1$) in the middle plot. Both clearly indicate collimating effect of the radiation, though the particles' non-zero angular momentum (given by the Keplerian value initially) naturally drives them somewhat off the axis. The top view projections onto the equatorial plane $(r\sin\theta\cos\phi,r\sin\theta\sin\phi)$ are shown in the right column (the concentrated disc above / the spread-out one below) is characterized mainly by azimuthal motion.}
\label{from-equat-rest}
\end{figure*}

\begin{figure*}
\includegraphics[width=0.84\textwidth]{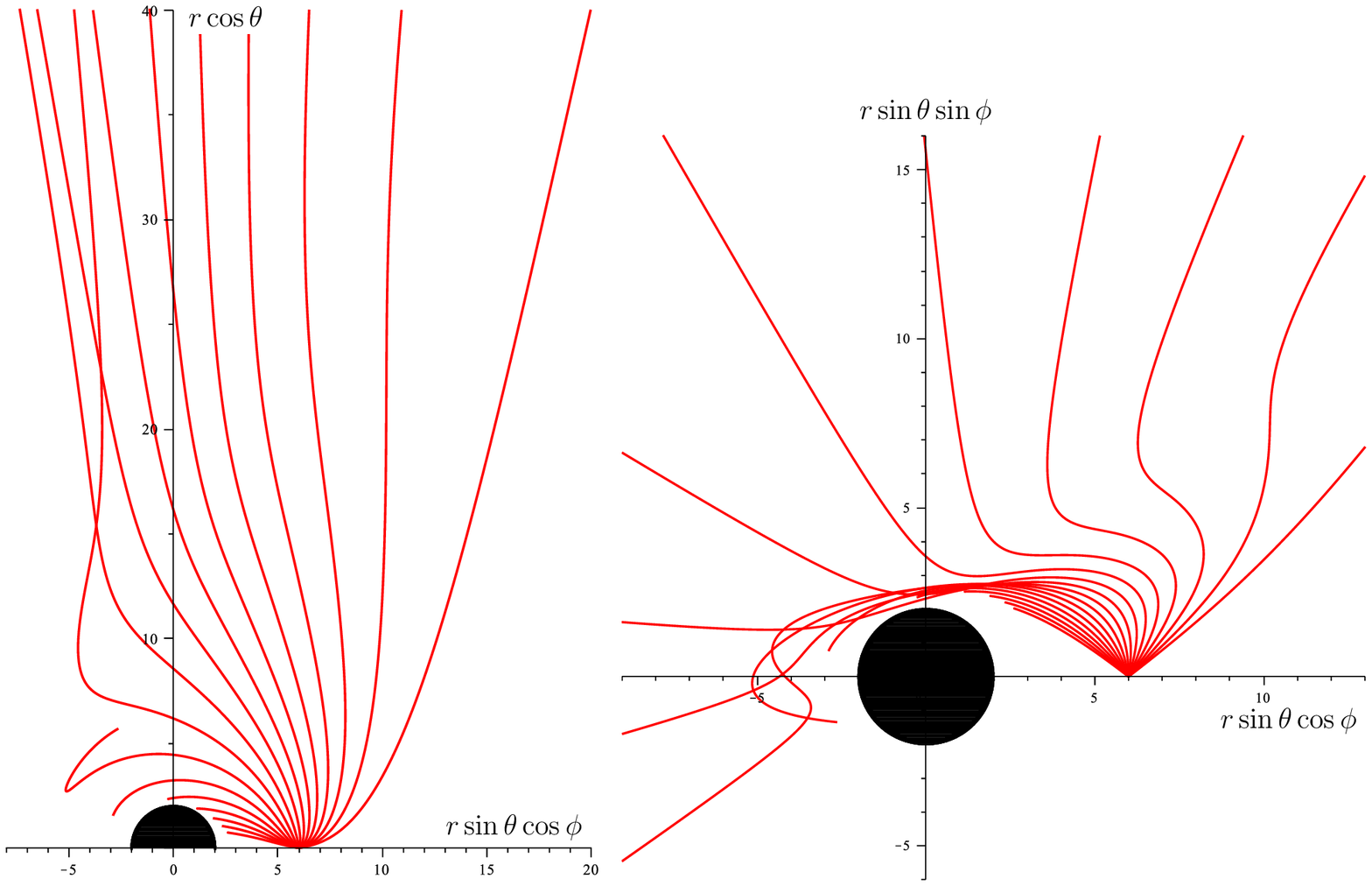}
\caption
{A fan of 17 particles released from $r=6M$ from the plane of the ``concentrated" disc (given by $\kappa=32$, $\lambda=2$). The initial velocity with respect to a local static observer has azimuthal component given by the Keplerian value, latitudinal component $-0.1$ (which translates into a slight push towards the axis) and radial component changing from $-0.8$, $-0.7$, $-0.6$, ..., to $+0.7$ and $+0.8$. Here the side view projection $(r\sin\theta\cos\phi,r\cos\theta)$ (not the meridional plane $r,\theta$-projection used above) is shown in the left plot and the top view in the right plot.}
\label{flux1-fanfrom6}
\end{figure*}

\begin{figure*}
\includegraphics[width=0.84\textwidth]{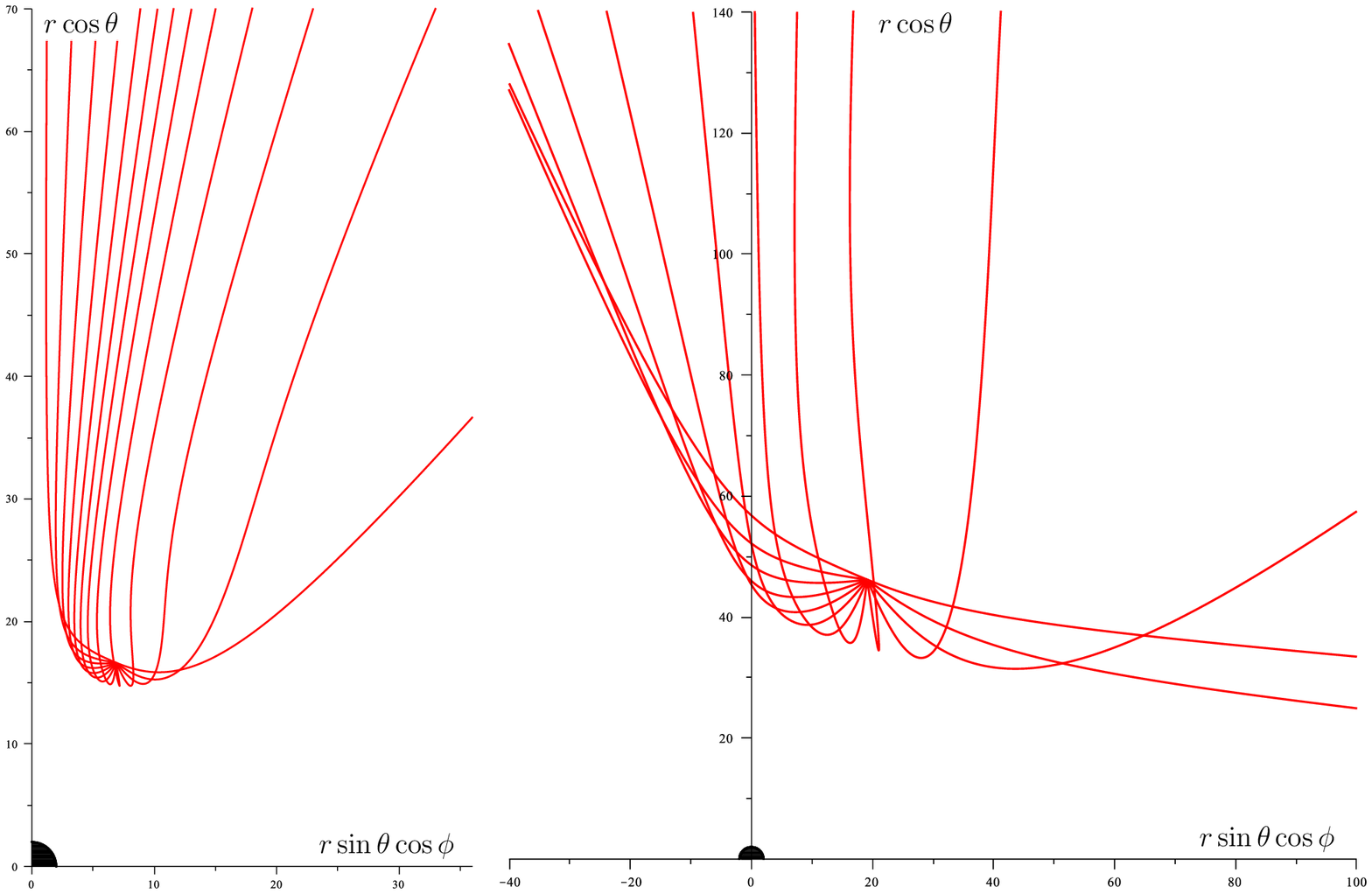}
\caption
{Side view of evolution of meridional fans of particles initially ingoing with respect to $r={\rm const}$, namely launched in 13 different directions towards the centre (by $15^{\circ}$): the initial velocity with respect to a local static observer has zero azimuthal component, while radial and latitudinal components are $-0.9\,(\sin\delta,\cos\delta)$ with $\delta=0$, $\pi/12$, $\pi/6$, $\pi/4$, \dots, $11\pi/12$, $\pi$. All the trajectories have no azimuthal motion, so the side view represents them completely. The {\it left plot} involves the ``concentrated" disc (given by $\kappa=32$, $\lambda=2$) and particles launched from $r=18M$, $\theta=\pi/8$, while the {\it right plot} involves the ``spread-out" disc ($\kappa=64$, $\lambda=1$) and particles launched from $r=50M$, $\theta=\pi/8$. Apparently the radiation ``blows" the particles away from the discs effectively and thus accelerates and collimates them (has them move into a relatively narrow cone with respect to which they would have filled otherwise).}
\label{ingoing-fans}
\end{figure*}

\begin{figure}
\begin{center}
\includegraphics[width=0.7\columnwidth]{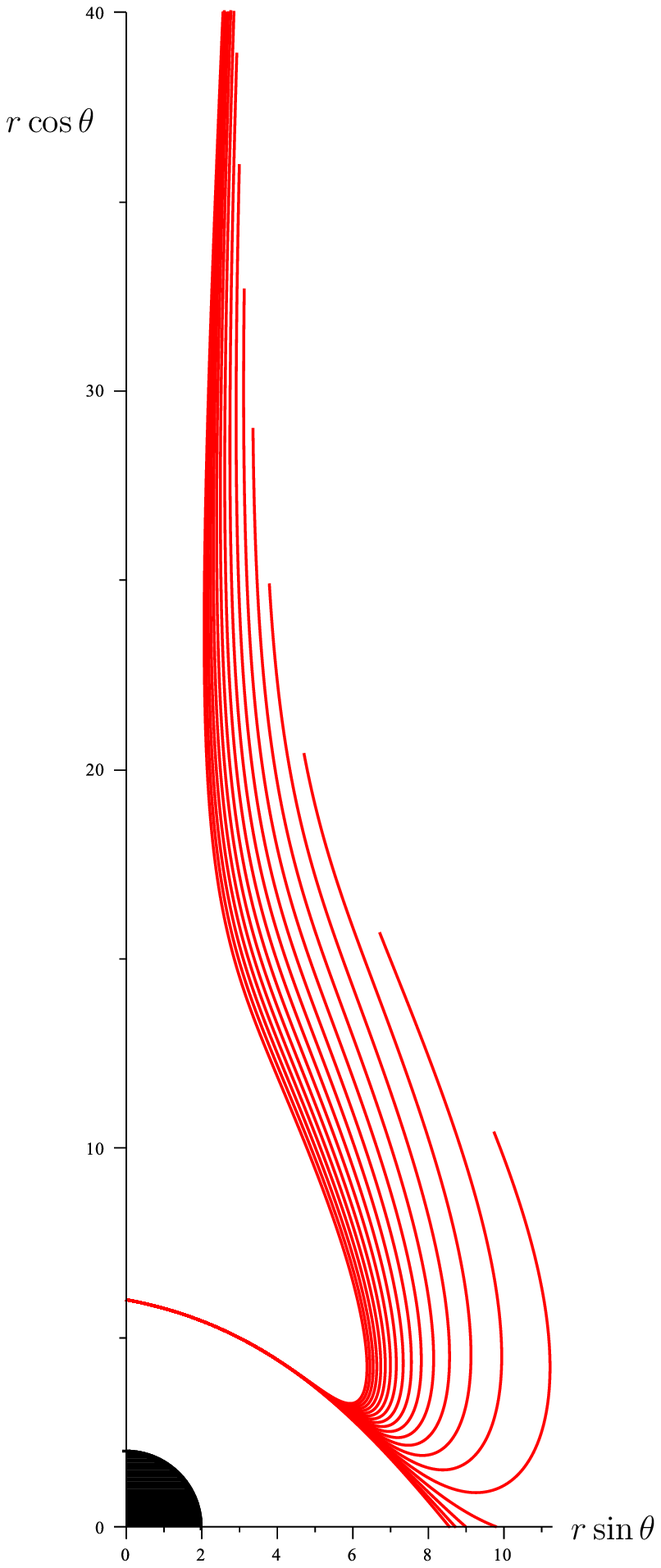}
\end{center}
\caption
{Effect of interaction strength, illustrated on a particle starting from $r=6M$ on the axis ($\theta=0$) with initial velocity $-0.2$ in the radial direction and $0.9$ in the latitudinal direction with respect to a local static observer. The trajectories differ in the value of $\tilde\sigma\,{\cal K}$, namely, going from the least bent to the most bent trajectories, $\tilde\sigma\,{\cal K}=1$, $2$, $3$, $4$, \dots, $20$. The rather concentrated flux is chosen ($\kappa=32$, $\lambda=2$). The particle starts from a region where the flux is relatively weak but then enters the region where it is almost maximal, which results in the bending of its trajectory, the more pronounced the stronger is the ``coupling" $\tilde\sigma\,{\cal K}$. The plot represents the trajectories completely, since they have no azimuthal component.}
\label{flux1-interaction-strength}
\end{figure}

\section{Examples of particle trajectories}
\label{examples}

The final step is to study the equations of motion (\ref{equation-of-motion}) numerically, in order to see whether and under which conditions the test particles tend to be accelerated and/or collimated along the symmetry axis.

\subsection{Choosing the scale factors}

First, one has to ``connect with nature" by choosing reasonably several free scale factors, namely the globally constant square ${\cal K}$ of the photon angular momentum, the constant multiplicative factor which scales the energy density of the flux $\Phi^2$, and the constant $\tilde{\sigma}$ which scales the efficiency of the photon $\!\rightarrow\!$ particle momentum transfer. As discussed in previous papers (see mainly \citealt{BiniJS-09}, last part of section 2), it is advantageous to follow \cite{Robertson-37} and combine all these factors into a single effective quantity (denoted by $A$) which has a useful interpretation. Namely, for a purely radially outgoing flux in a spherically symmetric field, it is given by
\[A=\tilde\sigma\,\Phi^2 E^2 r^2,\]
which is constant and equal to $M$ when the flux has exactly the Eddington value.
(The connection between the quantities $A$, $\tilde\sigma$ and luminosity was explained in \cite{BiniJS-09}, sections 3.1 and 3.2.)
Our flux is surely not radial and its photons do not have the same energy, so such a quantity is not constant in general and can only loosely be related to the Eddington luminosity, but it is still very helpful when trying to adjust the scheme to realistic parameters, with the $A=M$ value serving as ``benchmark".
Let us add that the actual accretion-disc flux may perhaps be highly super-Eddington, mainly if it comes out of the system in a direction where it does not counteract accretion; however, super-Eddington flux is not likely to be generated by a {\em thin} accretion disc.

More specifically, we have $E^2=\frac{\cal K}{b^2}=\frac{\cal K}{r_{\rm eq}^2}\left(1-\frac{2M}{r_{\rm eq}}\right)$, so in the equatorial plane the expression can be written in terms of ${\cal K}$ and $r_{\rm eq}$ alone,
\begin{equation}
  A=\tilde\sigma\,\Phi^2 {\cal K}\;\frac{r^2}{r_{\rm eq}^2}\left(1-\frac{2M}{r_{\rm eq}}\right)
   \stackrel{r=r_{\rm eq}}{\longrightarrow} \tilde\sigma\,\Phi^2 {\cal K}\left(1-\frac{2M}{r_{\rm eq}}\right).
\end{equation}
Since all components of the ``vertical"-photon four-momentum (\ref{photon-momentum}) are scaled by $\sqrt{\cal K}$, the force term in the equation of motion
\[
  -\tilde{\sigma}\,(\delta^\mu_\nu+u^\mu u_\nu)\,T^\nu{}_\lambda u^\lambda=
  -\tilde\sigma\,\Phi^2\,p_\nu p_\lambda\,u^\lambda(g^{\mu\nu}+u^\mu u^\nu)
\]
is proportional to $\tilde\sigma\,\Phi^2 {\cal K}$, which is well estimated by $A$ since the remaining factor $\left(1-\frac{2M}{r_{\rm eq}}\right)$ lies between $1/3$ and 1 and is typically close to 1.
Hence, for example, having the solution for $\Phi^2$ (namely the converging-flux solution $\Phi^2_-$), one can take its equatorial maximum ${\rm max}(\Phi_{\rm eq}^2)$ and then choose $\tilde\sigma\,{\cal K}$ according to $\tilde\sigma\,{\cal K}\sim\frac{A}{{\rm max}(\Phi_{\rm eq}^2)}\,$. For a ``10-times Eddington" disc one simply takes 10 times more.

\subsection{Choice of the approximation for photon trajectories}

The weak-field approximations---like that obtained by linearization in $M$---generally yield trajectories ``less bent about" the central gravitating body. When using such an approximation for our photon field, this means that both the trajectories of individual photons and the corresponding flux are oriented, at generic location, more in the vertical direction (they are less affected by the centre) than they would be in the exact description. For the flux this imperfection is no issue (our solutions are anyway represented by everywhere positive and smooth functions), but the approximate description of the individual trajectories can actually cause problems. Namely, since the centre's field is effectively weakened, an occurrence of a photon at a particular location may lead to inferring wrongly that it must have started from very close to (or even {\em below}) the horizon (a horizon is not actually present in an approximate description). In our case, the main problem occurs when the particle is close to the equatorial plane (especially if it is also at small radius), in particular, when one asks from where the photon started which should hit the particle there: according to the approximate picture, the photon which started from the opposite half of the equatorial plane can only get there if it started very close to the horizon or even from $r_{\rm eq}<2M$. Hence, in a certain region close to the equatorial plane the approximation is not usable for the ``secondary" photons (those which have already crossed the symmetry axis), because it would lead to negative $\left(1-\frac{2M}{r_{\rm eq}}\right)$ there and thus to imaginary $b$ and $E$. (This can be simply checked by plotting the formula (\ref{req,no-linearization}) for $r_{\rm eq}$ in the $\epsilon^\theta=+1$ case.)

There are two possible responses to this issue.
The first possibility is to take into account the secondary photons only outside the region where the above problem occurs. This is a reasonable option since the secondary flux is negligible anyway in the equatorial region close to the centre (while the primary flux is the strongest there, on the contrary).
The second possibility is to use a better approximation for the individual photon trajectories, without necessarily abandoning the flux density $\Phi^2$ obtained from the weak-field approximation. (We saw that the latter yields a well-behaved result.) Such an option might be considered inconsistent, yet still it is better than using the linear approximation ``consistently" (for the description of photon trajectories as well as for the continuity equation): the particle motion is mainly misrepresented if the impacting photon momenta are not correct, especially in the initial phase of motion close to the black hole (and remember that the momenta also enter the energy-momentum tensor), whereas details of the radiation density field are not that crucial; if the distortion due to approximation is not very large, one can understand the result as representing a field emitted by some slightly different source, which is no problem, because the radial profile of the radiation flux was chosen ``by hand" anyway (though of course in accord with predictions of the accretion-disc theory).

Therefore, in order to be able to also treat the innermost region close to the black hole properly, it is crucial to approximate the photon trajectories very accurately and, in addition, to be able to solve---at least in linear order in $M$, say---the form of the continuity equation obtained after substituting this approximation. As already discussed in section \ref{approximation}, we suggest and will use the approximation of the photon trajectories by the parametrized family of curves (\ref{our-approximation}) which can be inverted to yield the initial radius
\begin{align}
  r_{\rm eq} &=
  \frac{R_- +\sqrt{R_-^2+4Mr{\cal A}{\cal B}}}{2{\cal A}} \;, \qquad {\rm where} \\
  R_- &\equiv (r-\omega M)^2 (-\epsilon^\theta r\sin\theta+\alpha M)-Mr^2 \,, \nonumber \\
  {\cal A} &\equiv (r-\omega M)^2+Mr, \nonumber \\
  {\cal B} &\equiv 2r^2+\epsilon^\theta(r-\omega M)^2\alpha\sin\theta \,. \nonumber
\end{align}
Specifically, we have used this formula with the parameter values $\alpha=1.77$ and $\omega=1.45$ which yield a very accurate description, much better than the linearization in $M$ of the exact result and even better than the well known formula by Beloborodov which cannot be used below $r_{\rm eq}=4M$ and also is not easily invertible. (See Fig.~\ref{approximations} for comparison and \citealt{Semerak-15} for a more thorough account.) As already stressed above, it is favourable that for $\epsilon^\theta=-1$ and to linear order in $M$, this formula leads to the {\em same} continuity equation as the formula following from linearization in $M$ of the exact trajectory, so it is {\em consistent} with the flux $\Phi^2_-$ already found (at least up to linear order in $M$).

\begin{figure*}
\includegraphics[width=\textwidth]{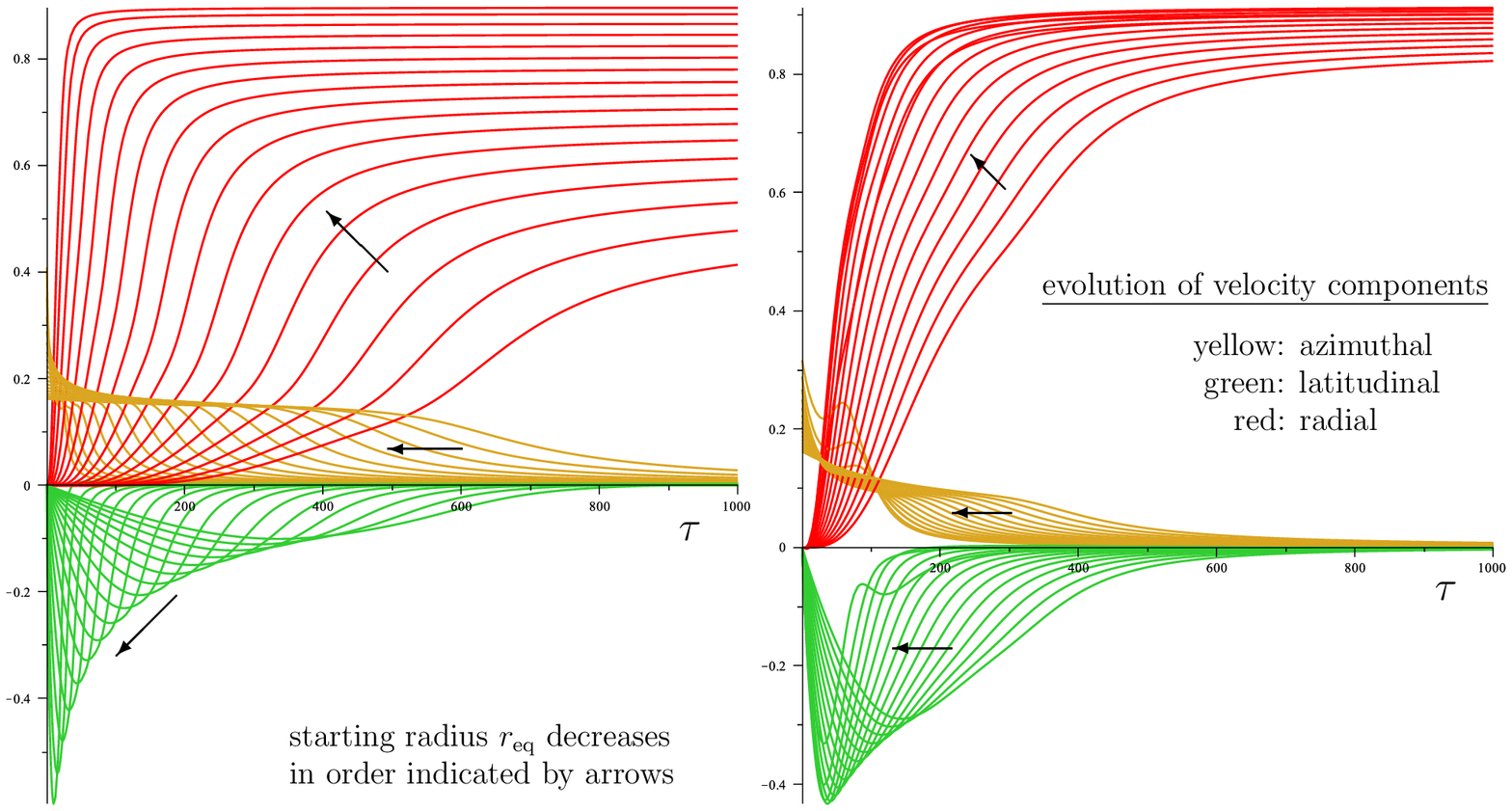}
\caption
{Time evolution of the physical velocity with respect to static observers at particle momentary locations, plotted for a set of particles starting from the radii $r=40M$, $38M$, $36M$, \dots, down to $8M$ (left) and $12M$ (right), from very near the equatorial plane, as driven by the more concentrated ($\kappa=32$, $\lambda=2$; left plot) and more spread-out radiation flux ($\kappa=64$, $\lambda=1$; right plot). The same motions have been considered in Fig.~\ref{from-equat-rest}. Evolution of all three components of the relative velocity in proper time (horizontal axis, given in units of $M$) is shown; the azimuthal component is plotted in yellow (it starts at the local Keplerian orbital value $\tilde{v}\equiv\tilde{v}^{\tilde{\phi}}=[(r/M)-2]^{-1/2}$ and then falls to zero overall), the latitudinal component is plotted in green (it starts from zero, quickly goes negative and after reaching a minimum it approaches zero gradually) and the radial component is plotted in red (it starts from zero and increases monotonously). The arrows help to identify the curves: the closer to the centre the particle started, the smaller / the smaller (less negative) / the larger is its azimuthal / latitudinal / radial velocity at $\tau=500M$. (The only exception are several particles released from smallest radii in the case of the less concentrated flux on the right: they start below maximum of the flux density and are influenced less there than those exposed to stronger flux at slightly larger radii.)
The plots show that the particles are strongly affected by radiation, but their terminal velocity (almost entirely radial) is not more than $0.9$ (of the speed of light).}
\label{velocity-evolution}
\end{figure*}

\subsection{Numerical results}

The effect of the accretion-disc radiation on particles initially floating somewhere around the inner part of the disc is illustrated on Figs.~\ref{from-equat-rest}--\ref{velocity-evolution}; some of them involve the rather concentrated radiation flux ($\kappa=32$ and $\lambda=2$), while some consider the flux spread-out to larger radii ($\kappa=64$ and $\lambda=1$). We mostly adjust the parameters to what we called the ``10-times Eddington" disc, i.e., we choose $\tilde\sigma\,{\cal K}\sim\frac{A}{{\rm max}(\Phi_{\rm eq}^2)}\,$ with $A\simeq 10M$. Specifically, we have set $\tilde\sigma\,{\cal K}=20$ for the concentrated flux and $\tilde\sigma\,{\cal K}=220$ for the less concentrated one.
The plots were drawn in Schwarzschild coordinates $(r,\theta,\phi)$, with initial velocities specified with respect to local static observers, i.e. those whose four-velocity is proportional to the time-like Killing vector $\partial x^\mu/\partial t$. The ``physical" (locally measured) components of these relative velocities $\tilde{v}^{\tilde{\imath}}$ are related to four-velocity $u^\mu$ by
$\tilde{v}^{\tilde{\imath}}=\sqrt{\frac{g_{ii}}{-g_{tt}}}\,\frac{u^i}{u^t}$
(no summation over $i$).

We first released a set of particles from very near the equatorial plane from radii $r=40M$, $38M$, $36M$, \dots, $10M$, endowing them only with Keplerian value of the azimuthal velocity, thus with physical velocity with respect to a local static observer given by $\sqrt{M/(r-2M)}\,$, but no initial velocity in the radial or latitudinal direction. Such particles should best approximate the motion of material within real accretion disc, and also reflect the effect of the disc's radiation without any prejudices on how the motion out of the disc should begin. The results are shown in Fig.~\ref{from-equat-rest} and indicate that the radiation quite strongly drives the particles in the axial direction. (However, since the particles have non-zero angular momentum, namely given by Keplerian value initially, their trajectories are somewhat deflected from the axis by centrifugal force.) Fig.~\ref{flux1-fanfrom6} contains a fan of particles released from $r=6M$ from the plane of the ``concentrated" disc. The particles differ in the value of the initial radial velocity and their trajectories again indicate vertical push by the radiation. Fig.~\ref{ingoing-fans} shows fans of particles launched towards the centre from $r=18M$ (for the concentrated disc) or $r=50M$ (for the spread-out disc), $\theta=\pi/8$, with various initial velocities covering the whole ``ingoing" half-space (with respect to $r={\rm const}$). The effect of a different interaction strength is revealed by Fig.~\ref{flux1-interaction-strength} where a test particle is bounced off the inner part of the disc the more the higher value of the coupling $\tilde\sigma\,{\cal K}$ one sets. Finally, Fig.~\ref{velocity-evolution} illustrates the time evolution of all three components of the particle's relative velocity with respect to a local static observer for both concentrated and spread-out radiation flux and for the same motions as followed in Fig.~\ref{from-equat-rest}.

As also specified in the figures captions, Fig.~\ref{from-equat-rest} (its left and middle panels) is using meridional-plane projection $(r\sin\theta,r\cos\theta)$ where azimuthal motion is suppressed completely (this component is revealed by top views in the right-hand panel), whereas in Fig.~\ref{flux1-fanfrom6} we use side-view projection $(r\sin\theta\cos\phi,r\cos\theta)$ where the line-of-sight component ($r\sin\theta\sin\phi$) of motion is suppressed (while the right-hand panel again brings top view along the axis). In Figs.~\ref{ingoing-fans} and \ref{flux1-interaction-strength} both the projections give the same result since the trajectories shown there have no azimuthal motion at all (zero angular momentum). Hence, while some of the intersections occurring in the plots are only seeming (namely those in Fig.~\ref{from-equat-rest}), the trajectories in Fig.~\ref{ingoing-fans} do really intersect due to the stronger ``repulsive" effect of radiation on particles which have approached the disc more closely.

\section{Conclusions, remarks and plans}

The picture of a black-hole thin accretion disc shining mainly in directions perpendicular to its plane has lead us to consider a radiation flux starting just perpendicular from the equatorial plane of a Schwarzschild field and to check how such a vertical flux affects test particles around the disc (which would otherwise follow geodesics of the background space-time). Numerical examples confirm that it can drive the particles effectively in motion along the axis accelerating and collimating them in that direction. However, for an astrophysically relevant range of flux and interaction-strength parameters, the acceleration of particles in itself is not enough to explain the highly relativistic energies observed in some jets emanating from black-hole sources; namely, we have observed ``terminal" Lorentz factors not much larger than 2 in our examples. This conclusion agrees with observations made in the literature (see mainly the references given in the Introduction) and seems to be rather robust with respect to a detailed profile of the flux, so it can be expected to also hold for more sophisticated models of disc emission with this same radiation-particle interaction mechanism.

Also known from the literature is another experience: when the flux is very strong, its detailed distribution is much more important for the trajectory than the particle's initial velocity. Actually, as already voiced in \cite{BisnovatyiKB-77}: ``The difference in the initial velocity\dots practically does not affect the results, since a proton acquires a velocity an order of magnitude higher\dots in a time much smaller than the orbital period [as being pushed by radiation], and the initial condition is rapidly forgotten." (On the contrary, initial {\em location} of the particle {\em is} of course important.) Our plots do not fully comply with such an experience: we considered relatively strong luminosities, yet the trajectories of particles launched from {\em the same} point (Figs.~\ref{flux1-fanfrom6}, \ref{ingoing-fans}, \ref{velocity-evolution}) clearly differ from each other according to their initial velocities.

We have mainly focused on meridional-plane projection of the motion in order to see the vertical effect of the flux, but it is worth to mention that the top views attached in Figs.~\ref{from-equat-rest} and \ref{flux1-fanfrom6} (cf. also Fig.~\ref{velocity-evolution}) reveal that the azimuthal motion is also far from trivial. As already pointed out at the end of section \ref{interaction}, this is mainly due to the term $-\tilde\sigma u^\phi\,T_{\nu\lambda}u^\nu u^\lambda$ in the equation of motion which is non-zero in spite of the azimuthal symmetry of the gravitational background as well as of the radiation flux. (Let us once more refer to \cite{KoutsantoniouC-14} who focused just on the azimuthal effect and drew interesting conclusions for the disc's inner edge.)

One should mainly investigate now how the results would be modified by a more appropriate description of the radiation-particle interaction. In fact the inner parts of accretion discs mainly emit in the X-band (10--1000 keV, say) where one should incorporate Compton scattering (which is described by the Klein-Nishina cross-section in the rest frame of the particle) rather than resort to the Thomson-like limit where the interaction is only characterized by an effective ``coupling coefficient" $\tilde\sigma$ independent of frequency. The results by \cite{KeaneBS-01} who compared these two descriptions in the case of a relativistic spherical source would be important in such an advancement. Another possible improvement would be to proceed to a hydrodynamical description of matter. Needless to say, since our study is purely particle-like, the figures do not in general say how a blob of plasma would move above an accretion disc; such a question would have to be solved by a hydrodynamic or MHD code. Though at least a qualitative agreement might be expected (cf. arguments given by \citealt{MishraK-14}), for a fluid the intersections would presumably lead to a formation of shocks, after which the fluid trajectories might differ significantly from the test-particle ones.

With a more appropriate model of the interaction, one might also proceed to a better model of the disc radiation: emission in all directions should be taken into account, not just the emission perpendicular to the disc, even though the ``vertical" pattern might represent a reasonable overall picture. Also, the radiation should correspond to that emitted by {\em orbiting} matter, so generically having some angular momentum. (However, due to the emission in all directions, there would also be present photons with zero angular momentum which can reach the symmetry axis.) Rotation should also be incorporated into the gravitational field, proceeding to the Kerr background. Finally, one should ensure that the innermost region is also endowed with a {\em ``correct"} flux, which would require, besides a very good description of the photon motion (we hope to have employed a very reasonable approximation here), to solve the continuity equation more accurately than up to linear order in $M$. We are confident that progress can be made along all these routes.

The last point we want to touch on is the question of particle escape. This has recently been treated by \cite{Stahl-KWA-13} and \cite{MishraK-14} with motivation to learn how changes of the centre's luminosity (spherically symmetric in their case) influence particle corona around, in particular, how strong burst is needed for a considerable coronal ejection. In our present paper, rather large luminosities have been chosen and all the particles whose trajectories are shown in the plots escaped to arbitrarily large distances, except one in the middle plot of Fig.~\ref{from-equat-rest} and seven in Fig.~\ref{flux1-fanfrom6}. However, these were all captured from the close vicinity of the black hole where all the above model imperfections are most serious. Before drawing more reliable implications about our system, specifically in case of moderate luminosities when details are even more important, one should proceed in the indicated directions.

\section*{Acknowledgements}

All authors thank ICRANet for support.
OS also thanks for support from Czech grant GACR-14-10625S.

\end{document}